\documentclass[lettersize,journal]{IEEEtran}
\usepackage{graphicx}                                           
\usepackage{subfigure}
\usepackage{booktabs}                                           
\usepackage{makecell}                                           
\usepackage{colortbl}                                           
\usepackage{multirow}                                           
\usepackage{cite}
\usepackage{amsmath}
\usepackage{amssymb}
\usepackage{amsthm}
\usepackage{pifont}

\usepackage{hyperref}
\usepackage{mathtools}
\usepackage{csquotes}
\usepackage{threeparttable}
\usepackage{subcaption}
\usepackage{xpatch}
\usepackage[inline]{enumitem}
\usepackage{url}
\usepackage{tabularx}
\usepackage{todonotes}
\usepackage{xcolor}
\usepackage{tikz}
\usepackage[edges]{forest}
\usepackage{balance}
\usetikzlibrary{mindmap,trees}
\usetikzlibrary{arrows,automata,shapes,positioning,shadows,trees}
\usetikzlibrary{shapes,snakes,shadows}
\usetikzlibrary{shapes.arrows}
\usetikzlibrary{calc,shapes,positioning}
\usetikzlibrary{decorations.text}
\usetikzlibrary{matrix,chains,positioning,decorations.pathreplacing,arrows}
\usetikzlibrary{shadows.blur}
\usetikzlibrary{shapes.geometric}
\usepackage[table]{xcolor}
\usepackage{graphicx}
\usepackage{longtable}
\usepackage{booktabs}
\usepackage{array}

\definecolor{SkillHeaderBlue}{RGB}{222,232,245}
\definecolor{SkillRowBlue}{RGB}{241,246,253}  
\definecolor{IconColor}{HTML}{3B82F6}       
\definecolor{StarColor}{HTML}{F59E0B}       
\definecolor{BtnBg}{HTML}{EFF6FF}           
\definecolor{BtnText}{HTML}{1D4ED8}         
\usepackage{fontawesome5}
\newcommand{\LinkGH}[1]{\href{#1}{\textcolor{black}{\faGithub}}}
\definecolor{hfyellow}{HTML}{FFD21E}
\definecolor{webblue}{HTML}{0000FF}
\newcommand{\LinkHF}[1]{\href{#1}{\textcolor{hfyellow}{\includegraphics[height=1em]{figures/hf-logo.png}}}}
\newcommand{\LinkWEB}[1]{\href{#1}{\includegraphics[height=1em]{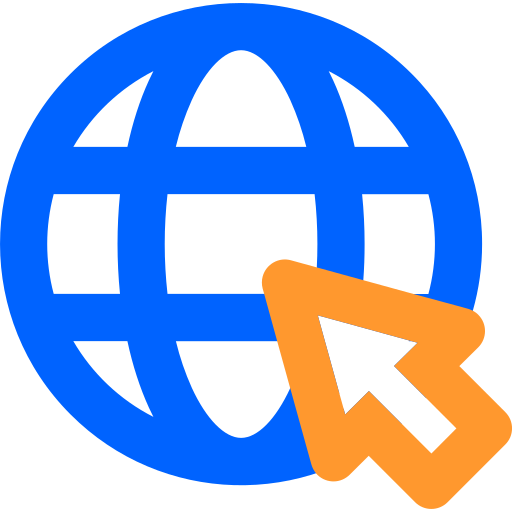}}}
\newtheorem{definition}{\bf Definition}

\definecolor{hidden-draw}{RGB}{20,68,106}
\definecolor{hidden-pink}{RGB}{255,245,247}
\definecolor{SkillHeaderBlue}{RGB}{224,236,248}
\definecolor{SkillRowBlue}{RGB}{244,248,252}
\tikzstyle{my-box}=[
    rectangle,
    draw=hidden-draw,
    rounded corners,
    align=left,
    text opacity=1,
    minimum height=1.5em,
    minimum width=5em,
    inner sep=2pt,
    fill opacity=.8,
    line width=0.8pt,
]

\tikzstyle{leaf-head}=[my-box, minimum height=1.5em,
    draw=gray!80,
    fill=gray!15,
    text=black, font=\normalsize,
    inner xsep=2pt,
    inner ysep=4pt,
    line width=0.8pt,
]
\tikzstyle{leaf-datasets}=[my-box, minimum height=1.5em,
    draw=red!70,
    fill=red!15,
    text=black, font=\normalsize,
    inner xsep=2pt,
    inner ysep=4pt,
    line width=0.8pt,
]
\tikzstyle{leaf-methods}=[my-box, minimum height=1.5em,
    draw=cyan!70,
    fill=cyan!15,
    text=black, font=\normalsize,
    inner xsep=2pt,
    inner ysep=4pt,
    line width=0.8pt,
]
\tikzstyle{leaf-metrics}=[my-box, minimum height=1.5em,
    draw=orange!80,
    fill=orange!15,
    text=black, font=\normalsize,
    inner xsep=2pt,
    inner ysep=4pt,
    line width=0.8pt,
]
\tikzstyle{modelnode-datasets}=[my-box, minimum height=1.5em,
    draw=red!80,
    fill=white,
    text=black, font=\normalsize,
    inner xsep=2pt,
    inner ysep=4pt,
    line width=0.8pt,
]
\tikzstyle{modelnode-methods}=[my-box, minimum height=1.5em,
    draw=cyan!100,
    fill=white,
    text=black, font=\normalsize,
    inner xsep=2pt,
    inner ysep=4pt,
    line width=0.8pt,
]
\tikzstyle{modelnode-metrics}=[my-box, minimum height=1.5em,
    draw=orange!100,
    fill=white,
    text=black, font=\normalsize,
    inner xsep=2pt,
    inner ysep=4pt,
    line width=0.8pt,
]

\AtBeginDocument{%
  }

\begin{document}

\title{A Comprehensive Survey on Agent Skills: Taxonomy, Techniques, and Applications}


\author{
\IEEEauthorblockN{
    Yingli Zhou,
    Shu Wang,
    Yaodong Su,
    Wenchuan Du,
    Yixiang Fang~\IEEEmembership{Member,~IEEE},
     Xuemin Lin~\IEEEmembership{Fellow,~IEEE} 
}

\IEEEauthorblockA{The Chinese University of Hong Kong, Shenzhen}

\IEEEauthorblockA{yinglizhou@link.cuhk.edu.cn; shuwang3@link.cuhk.edu.cn; yaodongsu@link.cuhk.edu.cn; 
doorvant@gmail.com;
fangyixiang@cuhk.edu.cn; 
linxuemin@cuhk.edu.cn;}

}



\maketitle

\begin{abstract}
Large language model (LLM)-based agents that reason, plan, and act
through tools, memory, and structured interaction are emerging as a
promising paradigm for automating complex workflows. 
Recent systems such as
OpenClaw and Claude Code exemplify a broader shift from passive response
generation to action-oriented task execution. Yet as agents move toward
open-ended, real-world deployment, relying on from-scratch reasoning and
low-level tool calls for every task become increasingly inefficient,
error-prone, and hard to maintain.
This survey examines this challenge through the lens of \emph{agent skills},
which we define as reusable procedural artifacts that coordinate tools,
memory, and runtime context under task-specific constraints. Under this view,
agents and skills play complementary roles: agents handle high-level
reasoning and planning, while skills form the operational layer that enables
reliable, reusable, and composable execution. Skills are therefore central
to the scalability, robustness, and maintainability of modern agent systems.
We organize the literature around four stages of the agent skill lifecycle
---representation, acquisition, retrieval, and evolution---and review representative methods, ecosystem resources, and application settings across
each stage. 
We conclude by discussing open challenges in quality control,
interoperability, safe updating, and long-term capability management.
All related resources, including research papers, open-source data, and projects, are collected for the community in \textcolor{blue}{\url{https://github.com/JayLZhou/Awesome-Agent-Skills}}.
\end{abstract}


\section{Introduction}
\label{sec:intro}

Large language model (LLM)-based agents are emerging as a powerful paradigm for automating complex tasks. Fundamentally, an LLM-based agent is an autonomous system that leverages an LLM as its cognitive engine to perceive its environment, interpret task context, reason over abstract goals, and execute actions through planning, tool use, memory retrieval, and structured interaction~\cite{brown2020gpt3,openai2023gpt4,ouyang2022instructgpt,react2022,shen2023hugginggpt,hong2024metagpt}. Recent pioneering systems, such as OpenClaw~\cite{openclaw2026} Manus~\cite{manus2026docs}, and Claude Code~\cite{anthropic2026claudecode},  vividly exemplify this paradigm, marking a broader transition in intelligent systems from passive response generation to proactive, action-oriented task execution.

As LLM-based agents are deployed in a growing range of scenarios and entrusted with increasingly complex tasks, tool augmentation has become a central design principle, enabled by APIs, plugins, and protocol layers such as MCP~\cite{anthropic2024mcp,openai2023functioncalling}. However, practical experience shows that mere access to tools does not determine \emph{when} a capability should be invoked, \emph{how} multiple tools should be coordinated, \emph{how} failures should be handled, or \emph{how} outputs should be validated. As tasks become more long-horizon and heterogeneous, relying on an agent to deduce these procedural steps from scratch for every task leads to severe brittleness, high latency, and unreliability. This ``procedural gap'' has emerged as a major bottleneck.

This gap motivates a fundamental shift toward a skill-centric view of agent systems. In this survey, we define \emph{agent skills} as reusable procedural artifacts that encode the specific ``how-to'' knowledge for coordinating tools, memory, and runtime context under concrete constraints~\cite{wang2023voyager,cai2024toolmakers,qian2023creator}. Within this framework, agents and skills form a highly synergistic \emph{hierarchical relationship}: the agent acts as the high-level cognitive planner responsible for intent interpretation and goal decomposition, while skills constitute the vital operational layer that translates these abstract plans into robust low-level execution. The importance of skills lies in their role as the agent's ``muscle memory.'' By externalizing procedural know-how into reusable artifacts, skills empower agents to bypass redundant step-by-step reasoning, drastically reduce execution errors, and transform transient actions into persistent capabilities that are easily retrieved, composed, revised, and governed across repeated tasks.

\begin{figure*}[t]
\centering
\includegraphics[width=2\columnwidth]{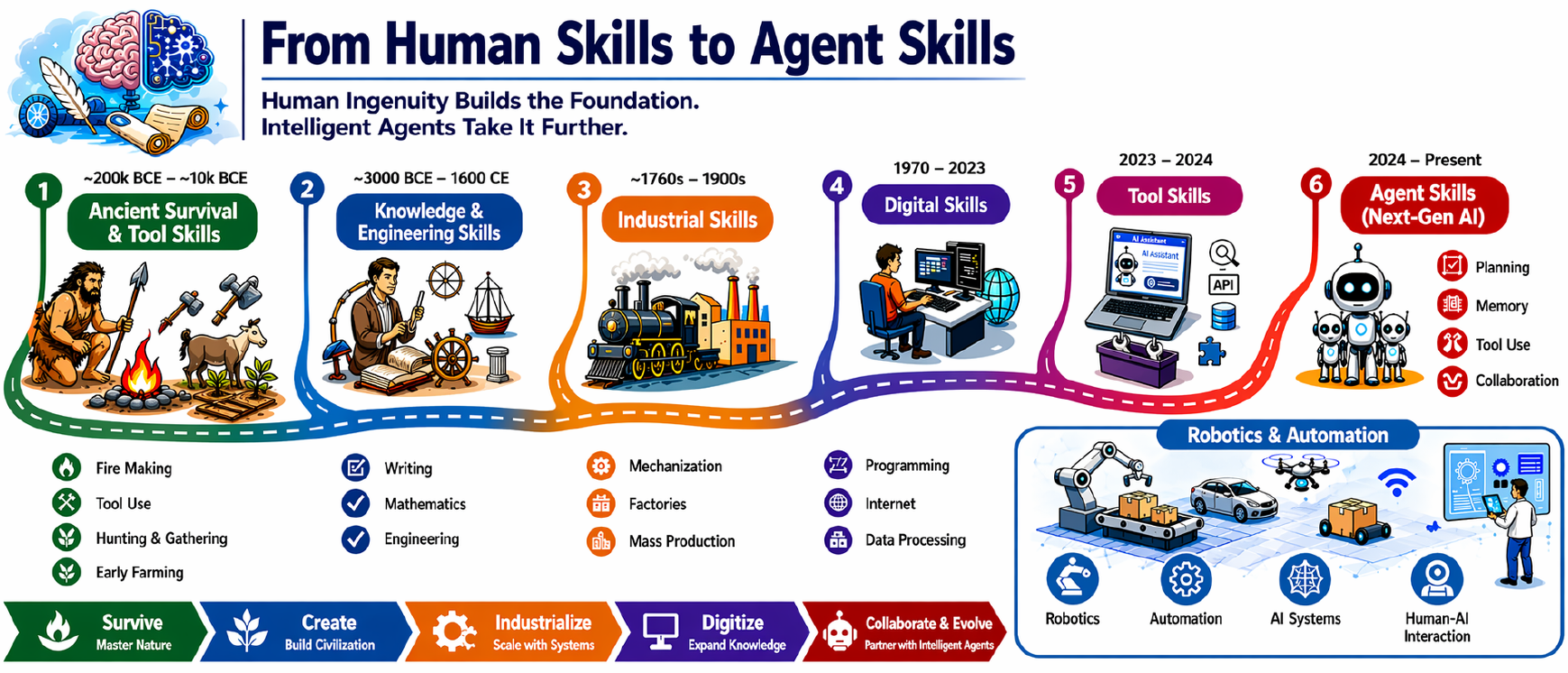}
\caption{Historical evolution of skills, from embodied human survival and craftsmanship to engineering, industrial, digital, and agent-era skill systems.}
\label{fig:skill-history}
\end{figure*}

More broadly, accumulating experience into reusable skills is a long-standing pattern in human learning. People do not solve every task from scratch; they progressively convert repeated practice, demonstrations, failures, and expert instruction into reusable procedures. As illustrated in Fig.~\ref{fig:skill-history}, this externalization process can be viewed as a long trajectory from embodied craft knowledge, to codified engineering procedures, to digital tools and programmable workflows, and now to agent-native skill ecosystems.
 
\begin{figure*}[t]
\centering
\includegraphics[width=1.9\columnwidth]{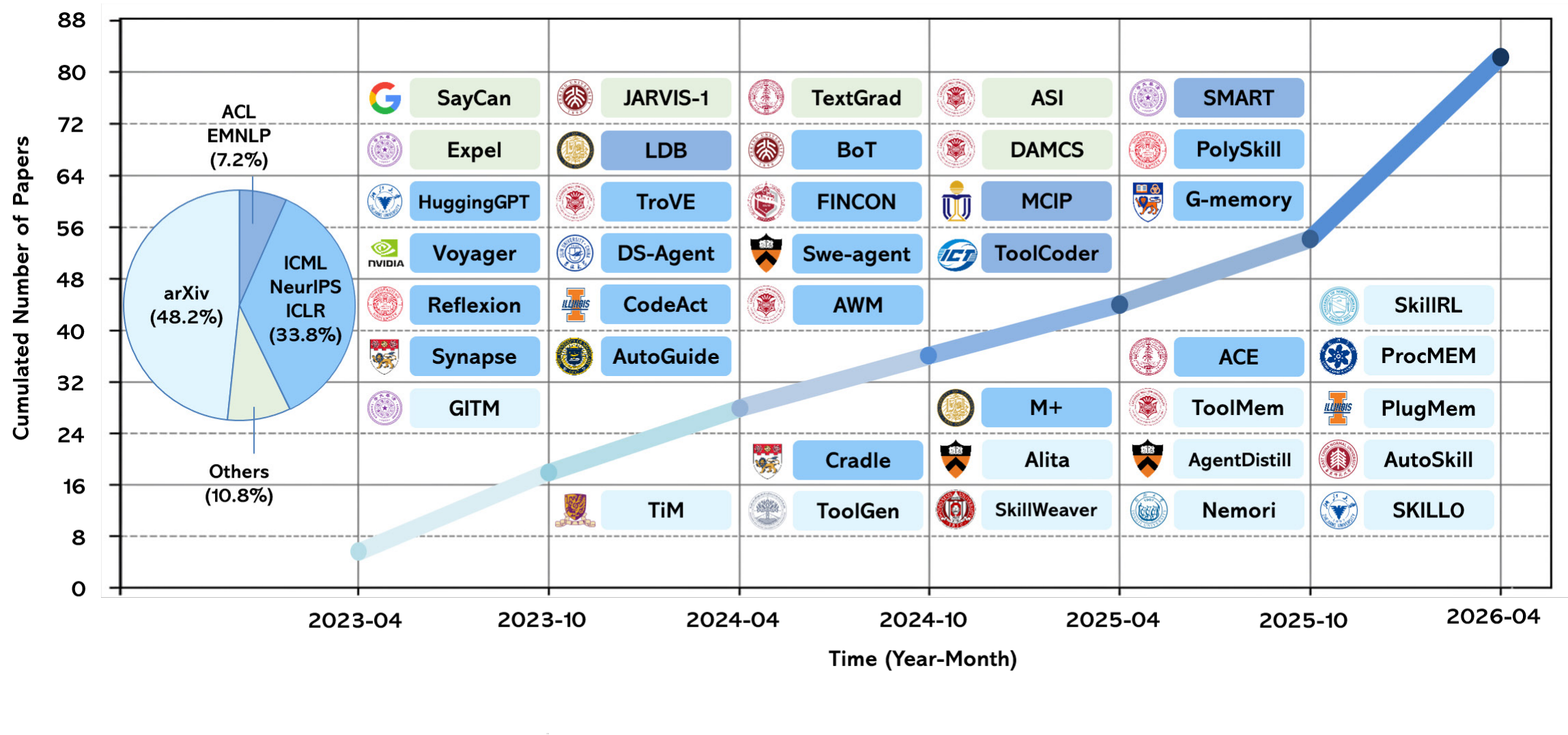}
\caption{Growth of research on agent skills from April 2023 to April 2026. The figure shows the cumulative number of representative papers together with example systems.}
\label{fig:paper-trend}
\end{figure*}

Against this backdrop, the recent LLM era has turned skill-centric agents into a fast-growing research area. Fig.~\ref{fig:paper-trend} shows the rapid growth of representative papers from April 2023 to April 2026. At the same time, progress is fragmented across multiple threads: skill acquisition from human expertise, traces, tasks, and corpora~\cite{wang2023voyager,cai2024toolmakers,qian2023creator}; retrieval from large and heterogeneous skill libraries~\cite{lewis2020rag,karpukhin2020dpr,du2024anytool}; runtime selection and composition under state and budget constraints~\cite{shen2023hugginggpt,hong2024metagpt,wu2023autogen}; and post-deployment revision, evolution, and governance~\cite{shinn2023reflexion,park2023generative}. This fragmentation motivates a focused and systematic survey of agent skills and skill-centric agent ecosystems.

To address this need, this survey provides a focused and systematic review of agent skills and skill-centric LLM agent ecosystems. We organize the literature around four lifecycle modules: skill representation, skill acquisition, skill retrieval, and skill evolution. Beyond method categorization, we review representative ecosystem resources, application settings, and synthesize open challenges in quality control, safety, cost, interoperability, maintenance, and long-term capability governance.

In summary, this survey makes the following contributions:
\begin{itemize}
 \item We identify agent skills as a foundational component of modern LLM agent ecosystems, explicitly characterizing their relationship with agents and articulating their pivotal role in bridging the procedural gap between raw tool access and robust task execution.
 
\item We organize existing research on agent skills around four core lifecycle stages ---skill representation, skill acquisition, skill retrieval, and skill evolution --- and review representative methods within each stage.

  \item We summarize representative agent skills platforms, application scenarios, and open challenges.
\item We outline promising research directions for agent skills.
\end{itemize}

The remainder of this paper is organized as follows. Section~\ref{sec:preliminaries} introduces core concepts and formal preliminaries. Section~\ref{sec:skill-taxonomy} presents the skill taxonomy used in this survey. Sections~\ref{sec:skill-acquisition}, \ref{sec:skill-retrieval-selection}, and \ref{sec:self-evolving-agents} discuss skill acquisition, retrieval, and evolution, respectively. Section~\ref{sec:related} discusses related work. Finally, Sections~\ref{sec:future} and \ref{sec:conclusion} outline future directions and conclude the paper.
\section{Preliminaries}
\label{sec:preliminaries}
\label{sec:background}

In this section, we introduce core preliminaries for skill-centric LLM agent ecosystems, including LLM-based agents, the procedural gap, the transition from tools to skills, and the lifecycle taxonomy used in this survey.

\subsection{LLM-Based Agents}
\label{sec:bg:agent}

An LLM-based agent is a system that uses a language model as its core reasoning engine to autonomously complete multi-step tasks through iterative interaction with an environment~\cite{react2022,park2023generative,wu2023autogen}. Its operation follows a perception--reasoning--action loop:
\begin{equation}
    o_t \;\rightarrow\; r_t \;\rightarrow\; a_t \;\rightarrow\; o_{t+1},
\end{equation}
where $o_t$ is the current observation, $r_t$ is the reasoning state, and $a_t$ is the resulting action. Recent systems such as OpenClaw and Manus exemplify this shift from passive response generation to action-oriented execution~\cite{openclaw2026,manus2026docs}. What separates modern agents from standalone LLMs is that they can \emph{act}: they query external systems, invoke tools, write and execute code, and coordinate with other agents~\cite{schick2023toolformer,shen2023hugginggpt,hong2024metagpt}. That action capability is the foundation of the capability stack studied in this survey.

\subsection{Agent Knowledge and the Procedural Gap}
\label{sec:bg:knowledge}

Agent behavior depends not only on the reasoning ability of the base model, but also on what knowledge is available when the agent must act. 
Here, we provide a simple distinction between passive and active knowledge to clarify where agent skills sit in the capability stack.

\textbf{Passive knowledge} is absorbed into the model's parameters before deployment through pre-training, supervised fine-tuning, and alignment procedures such as RLHF~\cite{brown2020gpt3,openai2023gpt4,ouyang2022instructgpt}. It includes factual associations as well as diffuse procedural priors such as instruction following, decomposition, and code or plan generation. These priors support general reasoning, but they remain static, opaque, and often weak in specialized or rapidly changing domains.

\textbf{Active knowledge} is obtained at runtime through interaction with the environment~\cite{lewis2020rag,schick2023toolformer,qin2023toolllm}. It includes retrieving documents, invoking APIs and tools, accessing MCP servers, executing externalized agent skills, and observing outcomes. Active knowledge is more dynamic and grounded than parametric knowledge, but access alone does not determine \emph{what} should be used, \emph{when} it should be used, \emph{how} it should be sequenced with other capabilities, or \emph{how} its outputs should be validated.

To bridge this procedural gap, agent skills package procedural knowledge into reusable procedural artifacts that can be stored, retrieved, revised, and governed across tasks. The rest of this survey therefore focuses on this active procedural layer rather than on agent knowledge as a whole.

\begin{figure*}[!th]
    \centering
    \vspace{-5mm}
    \resizebox{0.98\textwidth}{!}{
        \begin{forest}
            forked edges,
            for tree={
                grow=east,
                reversed=true,
                anchor=base west,
                parent anchor=east,
                child anchor=west,
                base=left,
                font=\normalsize,
                rectangle,
                draw=hidden-draw,
                rounded corners,
                align=left,
                minimum width=1em,
                edge+={darkgray, line width=1pt},
                s sep=3pt,
                inner xsep=0pt,
                inner ysep=3pt,
                line width=0.8pt,
                ver/.style={rotate=90, child anchor=north, parent anchor=south, anchor=center},
            },
            [
                Agent Skills, leaf-head, ver
                [
                    Skill Representation\\(\S\ref{sec:skill-taxonomy}), leaf-methods, text width=8.2em
                    [
                        Text-Based, leaf-methods, text width=7.0em
                        [Reflexion~\cite{shinn2023reflexion}{,} ExpeL~\cite{zhao2024expel}{,} BoT~\cite{yang2024buffer}{,} ReasoningBank~\cite{ouyang2025reasoningbank}{,} AWM~\cite{wang2024agentworkflowmemory}{,} Trace2Skill~\cite{ni2026trace2skill}{,} SayCan~\cite{saycan2022} {,}  DEPS~\cite{deps2023} {,} \\ Generative Agents~\cite{park2023generative}{,} GITM~\cite{zhu2023ghost}{,} RAP~\cite{hao2023reasoning}{,} Retroformer~\cite{yao2023retroformer}{,} MemGPT~\cite{packer2023memgpt}{,} TiM~\cite{liu2023thinkinmemory}{,} Self-Discover~\cite{zhou2024self}{,} \\ TextGrad~\cite{yuksekgonul2025optimizing}{,}  FINCON~\cite{yu2024fincon}{,} M+~\cite{wang2025mplus}{,} Learned Memory Bank~\cite{michelman2025enhancing}{,} Nemori~\cite{nemori2025}{,} Intrinsic Memory~\cite{intrinsicmemory2025}{,} SkillForge~\cite{mi2026procmem}, modelnode-methods, text width=50.7em]
                    ]
                    [
                        Code-Backed, leaf-methods, text width=7.0em
                        [Voyager~\cite{wang2023voyager}{,} SkillCraft~\cite{chen2026skillcraft}{,} PolySkill~\cite{polyskill2026}{,} ASI~\cite{inducingskills2025}{,} CUA-Skill~\cite{chen2026cua}{,}  MetaGPT~\cite{hong2024metagpt}{,} Eureka~\cite{ma2023eureka}{,} DS-Agent~\cite{yue2024dsagent}{,} \\ LDB~\cite{zhong2024ldb}{,}  CodeAct~\cite{wang2024codeact}{,} SWE-agent~\cite{yang2024sweagent}{,} ToolCoder~\cite{zhang2023toolcoder}{,} PSN~\cite{shi2026psn}, modelnode-methods, text width=50.7em]
                    ]
                    [
                        Hybrid-Based, leaf-methods, text width=7.0em
                        [JARVIS-1~\cite{wang2023jarvis1}{,} Synapse~\cite{zheng2024synapse}{,} SkillWeaver~\cite{skillweaver2025}{,} AgentSkillOS~\cite{li2026agentskillos}{,}  TPTU~\cite{ruan2023tptu}{,} talker-reasoner~\cite{christakopoulou2024agents}{,} DAMCS~\cite{damcs2025}{,} \\ GraphSkill~\cite{graphskill2026}{,} Alita~\cite{alita2025}, modelnode-methods, text width=50.7em]
                    ]
                ]
                [
                    Skill Acquisition\\(\S\ref{sec:skill-acquisition}), leaf-datasets, text width=8.2em
                    [
                        Human-Derived, leaf-datasets, text width=8em
                        [SkillNet~\cite{skillnet2026}{,} AgentSkillOS~\cite{li2026agentskillos}{,}  Agentic Skills~\cite{sokagenticskills2026}{,} SkillOS~\cite{skillos2026}{,} Agent Hospital~\cite{li2024agenthospital}, modelnode-datasets, text width=50.7em]
                    ]
                    [
                        Experience-Derived, leaf-datasets, text width=8em
                        [Voyager~\cite{wang2023voyager}{,} SkillCraft~\cite{chen2026skillcraft}{,} Reflexion~\cite{shinn2023reflexion}{,} ExpeL~\cite{zhao2024expel}{,} BoT~\cite{yang2024buffer}{,} Trace2Skill~\cite{ni2026trace2skill}{,} EverMemOS~\cite{hu2026evermemos}{,} HyperMem~\cite{yue2026hypermem}{,} \\ AWM~\cite{wang2024agentworkflowmemory}{,} Synapse~\cite{zheng2024synapse}{,} PolySkill~\cite{polyskill2026}{,} GITM~\cite{zhu2023ghost}{,} Retroformer~\cite{yao2023retroformer}{,} MemGPT~\cite{packer2023memgpt}{,}  Eureka~\cite{ma2023eureka}{,} TiM~\cite{liu2023thinkinmemory}{,}  M+~\cite{wang2025mplus}{,} \\ Learned Memory Bank~\cite{michelman2025enhancing}{,} G-Memory~\cite{zhang2506g}{,} Nemori~\cite{nemori2025}{,} AgentEvolver~\cite{agentevolver2025}{,} STULIFE~\cite{stulife2025}{,}  AutoRefine~\cite{qiu2026autorefine}{,} \\ ProcMEM~\cite{mi2026procmem}{,} SkillForge~\cite{mi2026procmem}, 
                        modelnode-datasets, text width=50.7em]
                    ]
                    [
                        Task-Derived, leaf-datasets, text width=8em
                        [CREATOR~\cite{qian2023creator}{,} ToolMakers~\cite{cai2024toolmakers}{,} Cradle~\cite{tan2024cradle}{,} CodeAct~\cite{wang2024codeact}{,} 
                     SkillWeaver~\cite{skillweaver2025}{,} SayCan~\cite{saycan2022}{,} ReAct~\cite{react2022}{,} DEPS~\cite{deps2023}{,} \\ RAP~\cite{hao2023reasoning}{,} MetaGPT~\cite{hong2024metagpt}{,} Self-Discover~\cite{zhou2024self}{,} LDB~\cite{zhong2024ldb}{,} SWE-agent~\cite{yang2024sweagent}{,} Alita~\cite{alita2025}, modelnode-datasets, text width=50.7em]
                    ]
                    [
                        Corpus-Derived, leaf-datasets, text width=8em
                        [AppAgent~\cite{zhang2023appagent}{,} AutoGuide~\cite{fu2024autoguide}{,} HuggingGPT~\cite{shen2023hugginggpt}{,} ToolLLM~\cite{qin2023toolllm}{,} WebArena~\cite{zhou2024webarena}{,} TPTU~\cite{ruan2023tptu}{,} ToolCoder~\cite{zhang2023toolcoder}{,} \\
                       DS-Agent~\cite{yue2024dsagent}{,} Corpus2Skill\cite{sun2026dontretrieve}{,} AgentDistill~\cite{qiu2025agentdistill}, modelnode-datasets, text width=50.7em]
                    ]
                ]
                [
                    Retrieval \&\\Selection (\S\ref{sec:skill-retrieval-selection}), leaf-metrics, text width=6em
                    [
                        Skill\\Retrieval, leaf-metrics, text width=4.8em
                        [
                            Dense Embedding, leaf-metrics, text width=8.5em
                            [Voyager~\cite{wang2023voyager}{,} SAGE~\cite{wang2025reinforcement}{,} AutoSkill~\cite{autoskill2026}{,} MemSkill~\cite{memskill2026}{,} ExpeL~\cite{zhao2024expel}{,} ReasoningBank~\cite{ouyang2025reasoningbank}{,} DS-Agent~\cite{yue2024dsagent}, modelnode-metrics, text width=47em]
                        ]
                        [
                            Sparse \& Keyword, leaf-metrics, text width=8.5em
                            [SAGE~\cite{wang2025reinforcement}{,} SkillWeaver~\cite{skillweaver2025}{,} AutoSkill~\cite{autoskill2026}{,} Memento-Skills~\cite{mementoskills2026}{,} SkillNet~\cite{skillnet2026}, modelnode-metrics, text width=47em]
                        ]
                        [
                            Generative Retrieval, leaf-metrics, text width=8.5em
                            [ToolGen~\cite{wangtoolgen}{,} ToolLLM~\cite{qin2023toolllm}, modelnode-metrics, text width=47em]
                        ]
                        [
                            Structure-Aware, leaf-metrics, text width=8.5em
                            [
                                Hierarchical, leaf-metrics, text width=8em
                                [SkillRL~\cite{skillrl2026}{,} AgentSkillOS~\cite{li2026agentskillos}{,} TOOL-PLANNER~\cite{liu2024tool}{,} SkillNet~\cite{skillnet2026}{,} GraphSkill~\cite{graphskill2026}{,} \\ MemGPT~\cite{packer2023memgpt}{,} G-Memory~\cite{zhang2506g}{,} Corpus2Skill~\cite{sun2026dontretrieve},
                                modelnode-metrics, text width=37.3em]
                            ]
                            [
                                Dependency Graph, leaf-metrics, text width=8em
                                [SkillWeaver~\cite{skillweaver2025}{,} CUA-Skill~\cite{chen2026cua}{,} ToolExpNet~\cite{zhang2025toolexpnet}{,} PSN~\cite{shi2026psn}, modelnode-metrics, text width=37.3em]
                            ]
                        ]
                    ]
                    [
                        Skill\\Selection, leaf-metrics, text width=4.8em
                        [
                            Context-Aware, leaf-metrics, text width=8.5em
                            [AutoGuide~\cite{fu2024autoguide}{,} MemSkill~\cite{memskill2026}{,} Memento-Skills~\cite{mementoskills2026}{,} ToolMem~\cite{xiao2025toolmem}{,} PlugMem~\cite{yang2026plugmem}, modelnode-metrics, text width=47em]
                        ]
                        [
                            Skill Composition, leaf-metrics, text width=8.5em
                            [SkillWeaver~\cite{skillweaver2025}{,} AWM~\cite{wang2024agentworkflowmemory}{,} ASI~\cite{inducingskills2025}{,} AgentSkillOS~\cite{li2026agentskillos}{,} CUA-Skill~\cite{chen2026cua}{,} HuggingGPT~\cite{shen2023hugginggpt}{,} Self-Discover~\cite{zhou2024self}, modelnode-metrics, text width=47em]
                        ]
                        [
                            Cost\&Utility-Aware, leaf-metrics, text width=8.5em
                            [MemSkill~\cite{memskill2026}{,} Memento-Skills~\cite{mementoskills2026}{,} SkillOrchestra~\cite{wang2026skillorchestra}{,} SkillsBench~\cite{li2026skillsbench}, modelnode-metrics, text width=47em]
                        ]
                        [
                            Feedback-Driven, leaf-metrics, text width=8.5em
                            [SkillRL~\cite{skillrl2026}{,} CUA-Skill~\cite{chen2026cua}{,} ToolExpNet~\cite{zhang2025toolexpnet}{,} ExpeL~\cite{zhao2024expel}{,} SMART~\cite{qian2025smart}, modelnode-metrics, text width=47em]
                        ]
                    ]
                ]
                [
                Skill\\Evolution\\(\S\ref{sec:self-evolving-agents}), leaf-methods, text width=6em
                [
                    Skill Revision, leaf-methods, minimum width=9.2em, text centered
                    [EvoSkill~\cite{evoskill2026}{,} Memento-Skills~\cite{mementoskills2026}{,} AutoSkill~\cite{autoskill2026}{,} XSkill~\cite{xskill2026}, modelnode-methods, minimum width=46.5em, text width=46.5em]
                ]
                [
                    Skill Validation, leaf-methods, minimum width=9.2em, text centered
                    [SkillWeaver~\cite{skillweaver2025}{,} ASI~\cite{inducingskills2025}{,} TroVE~\cite{trove2024}{,} PSN~\cite{shi2026psn}{,} Audited Skill-Graph~\cite{auditedskillgraph2025}{,} CoEvoSkills~\cite{coevoskills2026}, modelnode-methods, minimum width=46.5em, text width=46.5em]
                ]
                [
                    Policy Coupling, leaf-methods, minimum width=9.2em, text centered
                    [SkillRL~\cite{skillrl2026}{,} ARISE~\cite{arise2026}, modelnode-methods, minimum width=46.5em, text width=46.5em]
                ]
                [
                    Repository Evolution, leaf-methods, minimum width=9.2em, text centered
                    [Uni-Skill~\cite{uniskill2026}{,} SkillX~\cite{skillx2026}{,} SkillNet~\cite{skillnet2026}{,} SkillClaw~\cite{skillclaw2026}, modelnode-methods, minimum width=46.5em, text width=46.5em]
                ]
                [
                    Runtime Governance, leaf-methods, minimum width=9.2em, text centered
                    [SkillRouter~\cite{skillrouter2026}{,} PoisonedSkills~\cite{poisonedskills2026}, modelnode-methods, minimum width=46.5em, text width=46.5em]
                ]
            ]
            ]
        \end{forest}
    }
    \caption{The taxonomy for agent skills in this survey.}
    \label{fig:taxonomy-second}
\end{figure*}
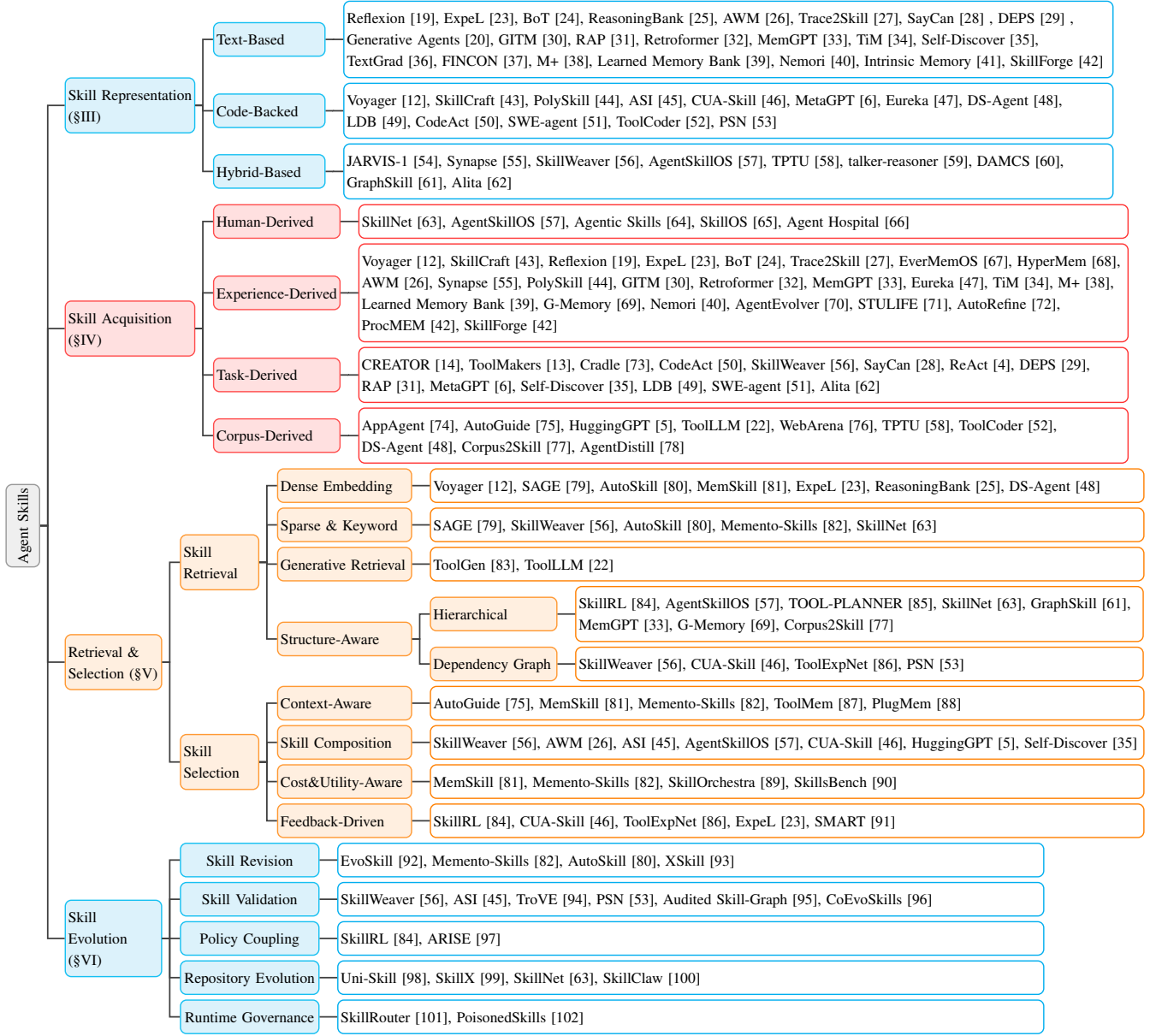

\begin{figure*}[]
\centering
\includegraphics[width=2.0\columnwidth]{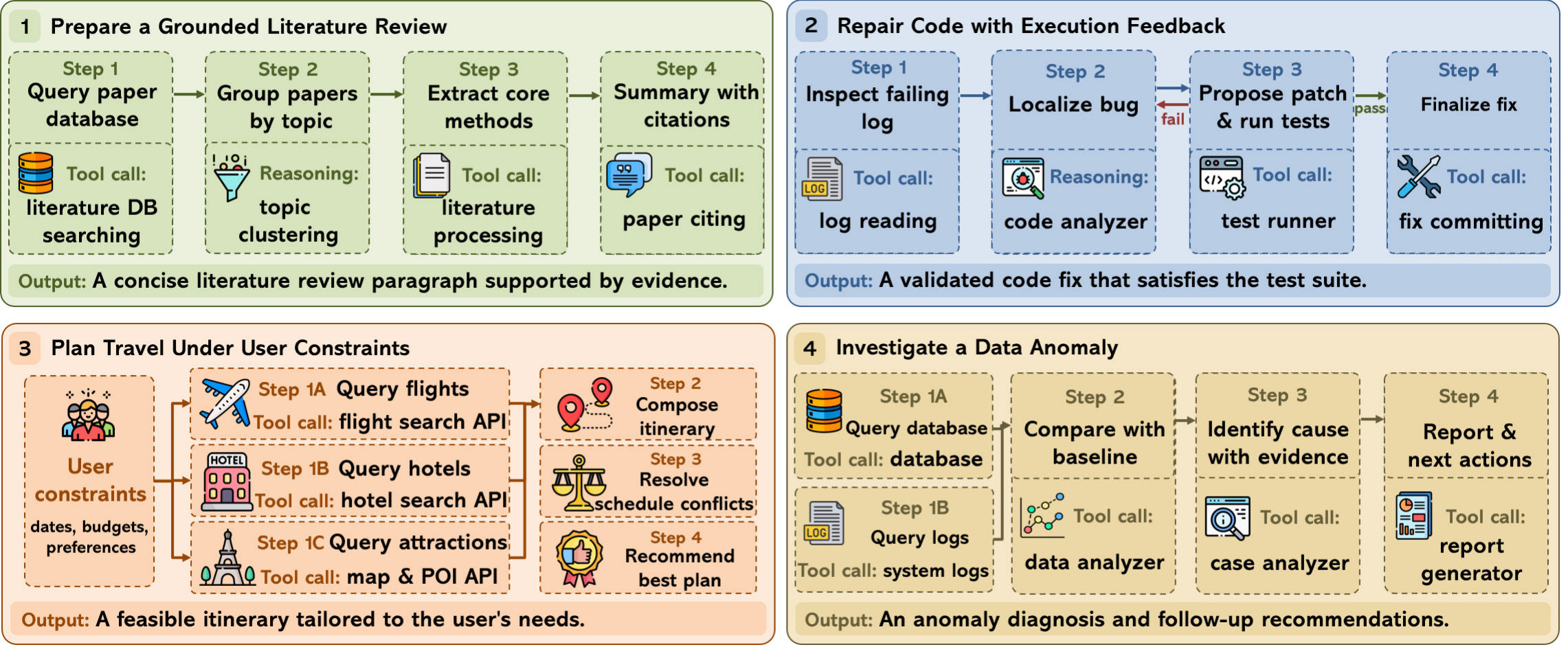}
\caption{Illustrative Examples of Agent Skills.}
\label{fig:skill_examples}
\end{figure*}

\subsection{From Tool Skills to Agent Skills}
\label{sec:bg:tools-to-skills}

The evolution of active LLM systems can be read as a transition from tool augmentation to skill augmentation. Tool-augmented agents made external capabilities practically usable at scale; skill-augmented agents emerged when researchers recognized that capability access alone was not enough, and that agents also needed reusable procedural artifacts for how those capabilities should be orchestrated in context.

Tools such as search engines, code interpreters, databases, and domain-specific APIs extend agents far beyond parametric memory alone~\cite{schick2023toolformer,qin2023toolllm,openai2023functioncalling,qin2024tool}. Standards such as the Model Context Protocol (MCP) further reduce integration friction by providing a unified mechanism for discovery and invocation across heterogeneous providers~\cite{anthropic2024mcp}. In effect, MCP makes it easier for an agent to \emph{reach} external capabilities.

But access does not by itself yield reliable behavior. A tool exposes an \emph{atomic capability}: it specifies \emph{what} can be done, not \emph{how} it should be used. A search tool does not say when search is preferable to memory retrieval; an API tool does not say what to do when the schema changes; a code interpreter does not say how outputs should be validated. MCP and similar infrastructures solve an \emph{interoperability} problem, not the \emph{procedural} problem of turning multiple tool calls into a robust workflow. As task complexity grows, that orchestration burden still falls back on the LLM at inference time and becomes a major source of brittleness~\cite{react2022,shinn2023reflexion}. This gap is addressed by introducing reusable procedural artifacts that specify when and how external capabilities should be applied.

\subsection{Skills: Definition and Formalization}

Rather than treating tool calls as isolated operations, systems store and reuse successful procedures as explicit \emph{skills}.

\begin{definition}[Agent Skill]
A skill is a reusable procedural artifact with bounded scope that externalizes
task-focused know-how: not only \emph{what} can be done, but \emph{when} to
act, \emph{how} to execute, what heuristics and failure modes matter, and how
to judge completion~\cite{ni2026trace2skill,ouyang2025reasoningbank}.
Formally, a skill is modeled as a tuple
\[
  S = (M,\;\mathcal{R},\;\mathcal{C}),
\]
where $M$ is the root instruction document that the agent can load and follow;
$\mathcal{R} = \{r_1,\ldots,r_K\}$ is a set of auxiliary resources---reference
documents, reusable templates, executable scripts, or domain artifacts that
extend what the skill accomplishes beyond $M$ alone~\cite{ni2026trace2skill};
and $\mathcal{C}$ encodes applicability conditions that govern when the skill
should be retrieved and applied, expressed as metadata, natural-language
descriptions, or embeddings~\cite{wang2023voyager,cai2024toolmakers,ni2026trace2skill}.
The tuple need not be fully instantiated in every system, but captures how
procedural knowledge becomes an explicit, inspectable object suitable for
storage, retrieval, and orchestration.
\end{definition}

Together, these components externalize task-focused know-how: not only \emph{what} can be done, but \emph{when} to act, \emph{how} to execute, what heuristics and failure modes matter, and how to judge progress and completion. The tuple need not be fully instantiated in every system, but captures how procedural knowledge becomes an explicit, partially inspectable object suitable for storage, retrieval, and orchestration. 
In practice, agent skills are increasingly managed through dedicated platforms such as SkillNet, ClawHub, SkillHub, SkillsMP, and Skills.sh (Table~\ref{tab:human_derived_examples}).

Skills differ from raw tools and MCP servers in that they encode \emph{situated} procedural knowledge (triggers, sequencing, fallbacks, pitfalls) and appear as \emph{bounded, reusable artifacts} that can be loaded, inspected, shared, and revised without becoming an undifferentiated handbook---a constraint that matters as libraries scale~\cite{wang2023voyager,cai2024toolmakers,chen2023skillit,ni2026trace2skill,ouyang2025reasoningbank,li2026agentskillos}. They need not be tool-centric: cognitive skills (e.g., review checklists, analysis workflows) mainly use internal knowledge but still supply structure and reuse beyond ad-hoc prompting. Tools expose operations; skills package know-how for using them in context.

\begin{table}[t]
    \centering
    \small
    \caption{Representative agent-skill platforms.}
    \label{tab:human_derived_examples}

    \setlength{\tabcolsep}{2pt}
    \renewcommand{\arraystretch}{1.08}
    \begin{tabular}{p{0.42\columnwidth} p{0.28\columnwidth} c}
            \toprule
            \rowcolor{SkillHeaderBlue}
            \textbf{Platform} & \textbf{Scale} & \textbf{Web} \\
            \midrule

            \rowcolor{SkillRowBlue}
            \textcolor{IconColor}{SkillNet} & \textcolor{StarColor}{300k+} & \LinkWEB{http://skillnet.openkg.cn/} \\

            ClawHub & \textcolor{StarColor}{40k+} & \LinkWEB{https://clawhub.ai/} \\

            \rowcolor{SkillRowBlue}
            \textcolor{IconColor}{SkillHub} & \textcolor{StarColor}{80k+} & \LinkWEB{https://www.skillhub.club/} \\

            SkillsMP & \textcolor{StarColor}{700k+} & \LinkWEB{https://skillsmp.com/} \\

            \rowcolor{SkillRowBlue}
            \textcolor{IconColor}{Skills.sh} & \textcolor{StarColor}{90k+} & \LinkWEB{https://skills.sh/} \\

            \bottomrule
    \end{tabular}
\end{table}

\subsection{The Ecosystem of Agent Skills}

Agent systems should be understood as ecosystems rather than isolated reasoning modules. Skills may be created from demonstrations, traces, documents, or feedback; indexed in repositories; retrieved and selected under task, latency, or budget constraints; executed with tools, memory, and other agents; and later revised, validated, or retired as environments change. Because these stages are tightly coupled, errors can propagate across the lifecycle: incomplete skills are harder to retrieve, weak retrieval signals can activate unsuitable procedures, and stale dependencies can break otherwise useful behavior.
In the following sections, we review each core lifecycle module in this ecosystem. As illustrated in Fig.~3, our taxonomy organizes existing studies from this lifecycle perspective.

\section{Skill Representation}
\label{sec:skill-taxonomy}

Building on Section~\ref{sec:preliminaries}, we classify how agent skills are \emph{packaged} in practice. Each skill consists of an instruction-based main document $M$, optional auxiliary resources $\mathcal{R}$, and trigger conditions $\mathcal{C}$. Semantically, skills externalize procedural knowledge, including operational structure, branching heuristics, and normative constraints~\cite{zhou2026externalization}, with $M$ serving as the primary human-readable representation. Although $M$ may take different forms, such as brief reminders, checklists, standard operating procedures (SOPs), or more detailed workflows, these variations are secondary to our taxonomy. The more consequential distinction across systems lies in how $\mathcal{R}$ is configured.


In practice, however, $M$ and $\mathcal{C}$ are often not sufficient on their own. Many skills rely on auxiliary resources $\mathcal{R}$ to provide domain references, reusable artifacts, or executable support, extending what the skill can accomplish beyond what is encoded in $M$ alone.

\subsection{Taxonomy by Resource Configuration}

We classify skills into three configurations based on the type of resources in $\mathcal{R}$: text-based, code-based, and hybrid.

\subsubsection{Text-backed skills} $\mathcal{R}$ consists of textual artifacts such as references, examples, templates, rubrics, or schemas. These resources improve grounding and reuse without introducing executable dependencies.

\subsubsection{Code-backed skills} $\mathcal{R}$ consists of executable artifacts such as scripts, helper functions, notebooks, or wrappers~\cite{wang2023voyager,cai2024toolmakers}. This enables repeatable subtasks and stronger operational determinism, but brings software packaging into the skill lifecycle: versioning, testing, and dependency management all become ongoing costs.

\subsubsection{Hybrid-resource skills} $\mathcal{R}$ combines textual and executable artifacts~\cite{anthropic2024mcp,ni2026trace2skill}, aiming to preserve interpretability while supporting deterministic execution. The coordination burden is highest here, as consistency must be maintained across documents, code, and their bindings.

\textbf{Example.} Fig.~\ref{fig:skill_examples} illustrates four representative examples of agent skills across different task scenarios. Each skill is organized as a reusable procedure composed of multiple steps, where each step may involve reasoning, tool invocation, or interaction with external resources. For example, a literature-review skill queries paper databases, groups papers by topic, extracts core methods, and produces a citation-supported summary. A code-repair skill inspects failing logs, localizes bugs, proposes patches, runs tests, and iteratively revises the fix based on execution feedback. Similarly, travel planning and anomaly investigation demonstrate how skills can coordinate parallel tool calls, satisfy user constraints, compare evidence, and generate actionable outputs.

\subsection{Comparison and Summary}

The three configurations differ not in the presence of $M$, but in what surrounds it. Text-based resources improve understanding; code-based resources improve execution reliability; hybrid configurations pursue both at the cost of greater complexity. The degenerate case—near-empty $\mathcal{R}$—places the full burden on $M$ alone. At ecosystem scale, this taxonomy shapes how skills are indexed, validated, maintained, and orchestrated~\cite{li2026agentskillos}, and provides a unified account of lightweight reminders, workflow packages, and richer skills with attached code and references.

\section{Skill Acquisition}
\label{sec:skill-acquisition}

Skill acquisition is the process of constructing or generating new skills. In practice, this means obtaining reusable procedural guidance or executable skill artifacts.

\begin{figure*}[h]
\centering
\includegraphics[width=2.1\columnwidth]{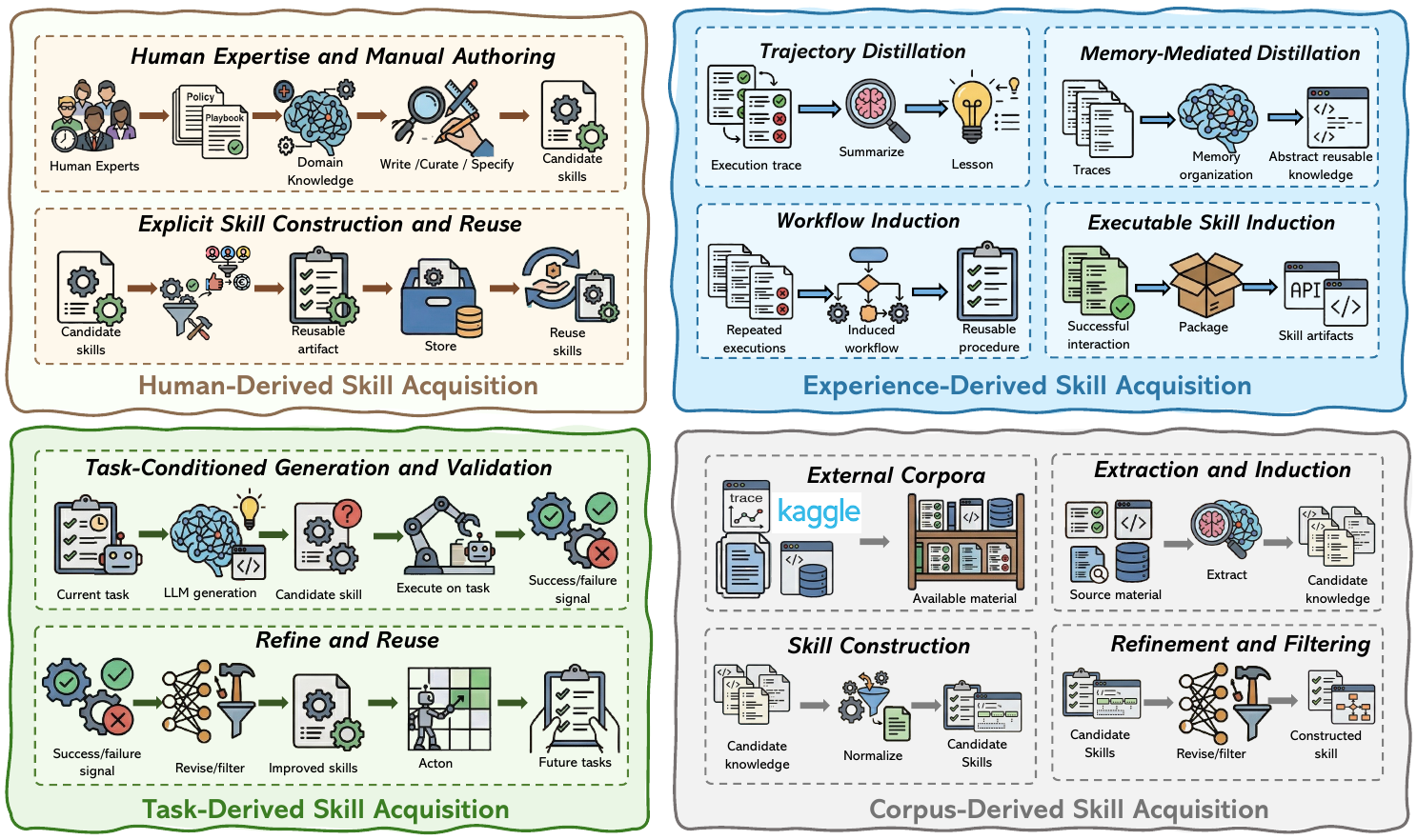}
\caption{Overview of skill acquisition methods.}
\label{fig:skill_obtain}
\end{figure*}
At a broad level, the literature can be organized into four families according to the dominant direct source from which a skill is obtained: {\Large \ding{182}}  human-derived acquisition, {\Large \ding{183}} experience-derived acquisition, {\Large \ding{184}} task-derived acquisition, and {\Large \ding{185}} corpus-derived acquisition. Human-derived acquisition obtains skills directly from expert knowledge and manual curation. Experience-derived acquisition builds them from trajectories, exemplars, or past executions. Task-derived acquisition constructs them on demand from the requirements of the current task. Corpus-derived acquisition extracts skills from external textual or structured corpora such as documents, repositories, and interface traces. 
Fig.~\ref{fig:skill_obtain} summarizes this broader landscape before the more focused discussion below.

\subsection{Human-Derived Acquisition}

Human-derived acquisition obtains skills directly from domain experts. In this group, people explicitly write reusable procedures, define their intended scope, and attach supporting materials or usage constraints when needed. 

Through repeated practice, they accumulate tacit know-how, exception-handling strategies, and context-sensitive judgment, which can then be externalized as reusable skills.
For example, doctors may summarize diagnostic experience into treatment procedures, medication prescriptions, and drug-specific usage recipes; engineers may codify troubleshooting workflows and operational playbooks; and policy experts may formalize review criteria and safety constraints. 
By doing so, human-derived acquisition therefore converts domain knowledge and professional experience into explicit procedural artifacts that agents can inspect, reuse, and adapt.

\begin{figure*}[]
\centering
\includegraphics[width=1.7\columnwidth]{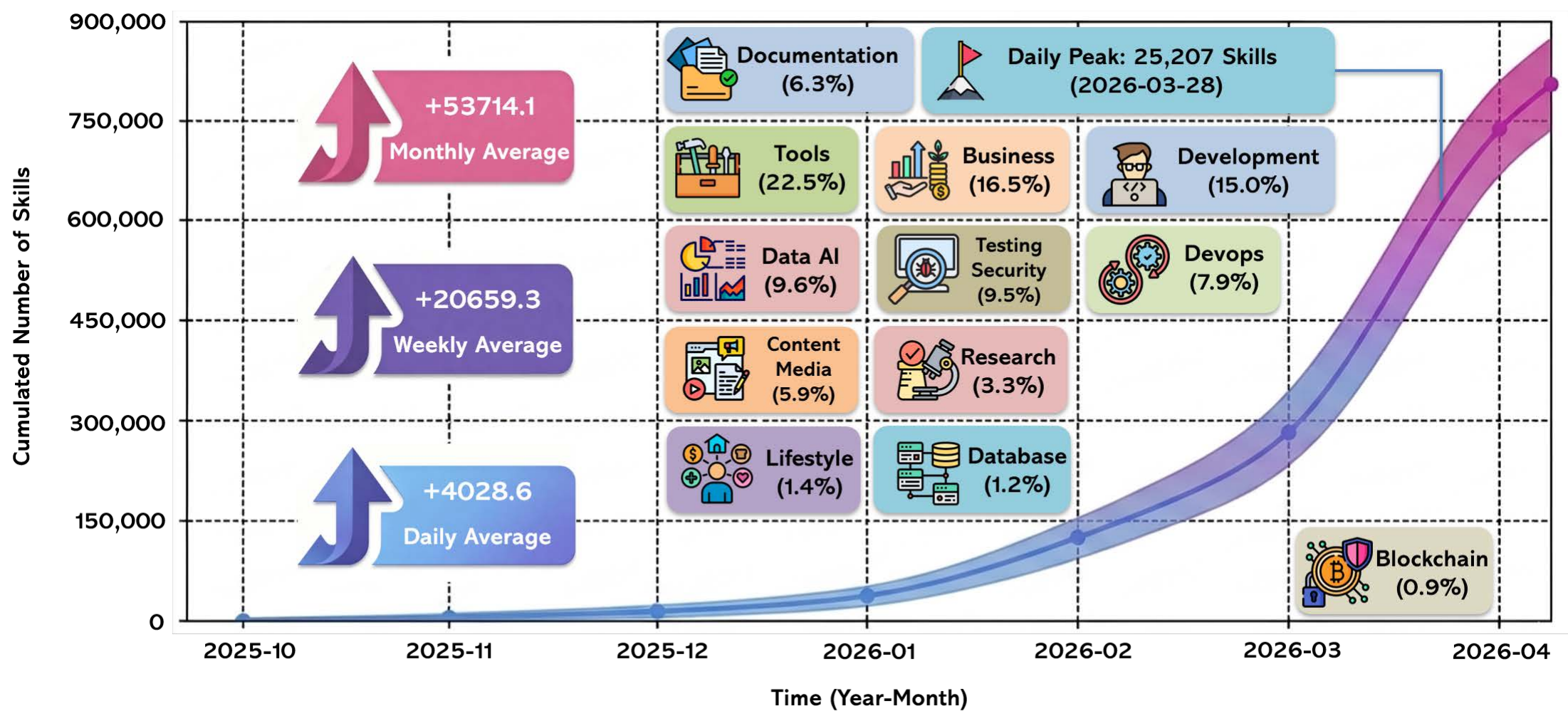}
\caption{The trend of cumulated number of human-derived skills over time.}
\label{fig:number_of_skills}
\end{figure*}

\textbf{Discussions.} The trade-offs of human-authored acquisition are straightforward. Its strength is precision: experts can encode tacit judgment, conventions, and safety-critical rules with fine semantic control. Its weakness is scalability, since manual curation is slow to grow and maintain; in practice, human-derived skills often serve as a seed layer for more automated acquisition.
Fig.~\ref{fig:number_of_skills} shows the recent growth and diversification of skills, based on statistics from SkillsMP (\url{https://skillsmp.com/}). This trend reflects a gradual shift: more skills designed by domain experts are being incorporated into agent platforms, as summarized in Table~\ref{tab:human_derived_examples}. Such expansion greatly improves agents' ability to perform specialized, domain-specific tasks.


\begin{table*}[]
\centering
\caption{Representative experience-derived methods organized by experience processing operators.}
\label{tab:acquisition-comparison}
\scriptsize
{\setlength{\tabcolsep}{3pt}
\renewcommand{\arraystretch}{1.18}
\resizebox{\textwidth}{!}{
\begin{tabular}{p{2.7cm} p{4.2cm} p{1.4cm} p{5.1cm} p{1.6cm}}
\toprule
\rowcolor{SkillHeaderBlue}
\textbf{Method} & \textbf{Source} & \textbf{Form} & \textbf{Key Characteristics} & \textbf{Venue/Year} \\
\midrule
\multicolumn{5}{c}{\textit{Selection}} \\
\midrule
Voyager~\cite{wang2023voyager} & Successful embodied trajectories & Code & Retains successful executable traces & NeurIPS'23 \\
\rowcolor{SkillRowBlue}
SkillCraft~\cite{chen2026skillcraft} & Successful tool trajectories & Code & Filters useful tool-use traces & arXiv'26 \\
\midrule
\multicolumn{5}{c}{\textit{Abstraction}} \\
\midrule
Reflexion~\cite{shinn2023reflexion} & Failed trajectories & Text & Converts failures into lessons & NeurIPS'23 \\
\rowcolor{SkillRowBlue}
ExpeL~\cite{zhao2024expel} & Successes and failures & Text & Abstracts reusable lessons & AAAI'24 \\
BoT~\cite{yang2024buffer} & Problem-solving traces & Text & Distills reasoning templates & NeurIPS'24 \\
\rowcolor{SkillRowBlue}
Trace2Skill~\cite{ni2026trace2skill} & Trajectories & Text & Consolidates trace-level skills & arXiv'26 \\
FINCON~\cite{yu2024fincon} & Decision episodes & Text & Distills decision insights & NeurIPS'24 \\
\rowcolor{SkillRowBlue}
AgentEvolver~\cite{agentevolver2025} & Processed trajectory sets & Text & Summarizes cross-task experience & arXiv'25 \\
\midrule
\multicolumn{5}{c}{\textit{Memory Organization}} \\
\midrule
TiM~\cite{liu2023thinkinmemory} & Interaction history & Text & Updates reflective memory & arXiv'23 \\
\rowcolor{SkillRowBlue}
G-Memory~\cite{zhang2506g} & Collective experience & Text & Builds hierarchical memory & NeurIPS'25 \\
Nemori~\cite{nemori2025} & Dialogue history & Text & Consolidates semantic memory & arXiv'25 \\
\rowcolor{SkillRowBlue}
Intrinsic Memory~\cite{intrinsicmemory2025} & Interaction history & Text & Maintains contextual memory & arXiv'25 \\
DAMCS~\cite{damcs2025} & Multimodal interactions & Hybrid & Organizes multimodal KG memory & arXiv'25 \\
\midrule
\multicolumn{5}{c}{\textit{Procedural Packaging}} \\
\midrule
\rowcolor{SkillRowBlue}
JARVIS-1~\cite{wang2023jarvis1} & Successful multimodal task executions & Hybrid & Retains task-plan procedures & TPAMI'25 \\
Synapse~\cite{zheng2024synapse} & Successful computer-control trajectories & Hybrid & Packages trajectory exemplars & ICLR'24 \\
\rowcolor{SkillRowBlue}
AWM~\cite{wang2024agentworkflowmemory} & Interaction traces & Text & Induces workflow skills & ICML'25 \\
PolySkill~\cite{polyskill2026} & Successful interaction traces & Code & Builds programmatic skills & ICLR'26 \\
\bottomrule
\end{tabular}
}}
\end{table*}

\subsection{Experience-Derived Acquisition}

Experience-derived acquisition treats an agent’s past runs—execution traces, exemplars, interaction histories, and feedback—as raw material from which it abstracts recurring patterns (including failures) into reusable, transferable skills, rather than leaving that information only as transient memory. This mirrors a familiar learning intuition—mine experience, distill procedure—and is currently the most heavily studied acquisition family, with the widest spread of concrete mechanisms and systems.

What chiefly differentiates methods is not the provenance of experience—typically traces, logs, and feedback from prior runs—but how that material is processed before it hardens into a skill. Concretely, systems may select and retain successful episodes as reusable exemplars; abstract long trajectories into compact lessons, heuristics, or declarative guidance; organize scattered evidence into structured memory that transfers across tasks; or package distilled know-how into richer artifacts such as workflows, APIs, or executable modules. Viewed this way, experience-derived acquisition is a pipeline from execution traces to skill artifacts, mediated by experience-processing operations (selection, abstraction/summarization, memory organization, and procedural packaging). These steps are seldom isolated: a single pipeline often chains several of them. The remainder of this section walks through each operation in turn.

A first common operation is \textbf{selection}. Since not all experience is equally useful for skill construction, systems often begin by filtering historical trajectories and retaining only those that are successful, informative, or representative.
Voyager is a canonical example: it keeps successful embodied trajectories and uses them as the basis for later executable skill construction~\cite{wang2023voyager}. SkillCraft follows a similar pattern in tool-use settings, where useful tool trajectories are selected before being further abstracted into executable skills~\cite{chen2026skillcraft}. In this sense, selection acts as an experience-quality control stage: it determines which parts of execution history become reusable raw material, while failed or noisy traces are either discarded or passed to abstraction-oriented methods as corrective evidence.

A second operation is \textbf{summarization and abstraction}. Here, concrete traces are compressed into reusable procedural knowledge such as lessons, heuristics, guidelines, or declarative skill descriptions. Reflexion extracts verbal reflections from failed attempts, turning specific errors into short corrective rules for future use~\cite{shinn2023reflexion}. ExpeL goes beyond single-trajectory reflection by abstracting higher-level lessons from accumulated successes and failures~\cite{zhao2024expel}. FINCON stores manager-level insights distilled from prior financial decision episodes as reusable guidance, while Buffer of Thoughts distills prior problem-solving experience into reusable reasoning templates that can later be instantiated on new tasks~\cite{yu2024fincon,yang2024buffer}. Think-in-Memory (TIM) likewise uses post-thinking to extract reusable natural-language experience, and Trace2Skill pushes this direction further by hierarchically consolidating local trajectory-level lessons into explicit reusable skill artifacts~\cite{liu2023thinkinmemory,ni2026trace2skill}. Across these methods, the shared pattern is that episodic executions are not preserved in full detail, but compressed into more compact knowledge units that are easier to retrieve and reuse.

A third operation is \textbf{memory organization}. Instead of treating each trajectory independently, some methods reorganize accumulated experience into more persistent and structured memory forms. Think-in-Memory integrates reflection into memory updates so that new executions reshape later reasoning rather than remaining as isolated records~\cite{liu2023thinkinmemory}. G-Memory makes this structuring more explicit by organizing collective experience into hierarchical memory graphs that support reuse at different abstraction levels~\cite{zhang2506g}. Nemori provides a related example in which episodic interactions are further distilled into more stable semantic memory, while Intrinsic Memory Agents maintain structured contextual memory so that accumulated experience can remain usable over longer horizons~\cite{nemori2025,intrinsicmemory2025}. AgentEvolver summarizes cross-task experience into reusable natural-language guidance units for later exploration, and DAMCS organizes multimodal past interactions into a hierarchical knowledge-graph memory that supports retrieval and structured communication for cooperative planning~\cite{agentevolver2025,damcs2025}. These systems differ in representation and scope, but all illustrate the same design move: past execution is reorganized into structured memory that remains available for later decision making and adaptation.

A fourth operation is \textbf{procedural induction and packaging}. Rather than extracting only textual guidance, some systems convert repeated successful executions into richer reusable procedures such as workflows, plans, APIs, or executable modules. Agent Workflow Memory is a representative case because it induces reusable workflows directly from prior interaction traces and stores them as workflow skills with triggers, multi-step procedures, and parameter slots~\cite{wang2024agentworkflowmemory}. Related systems such as JARVIS-1, Synapse, and Ghost in the Minecraft likewise preserve more structured procedural support from successful experience, often retaining ordering, decomposition, or action dependencies~\cite{wang2023jarvis1,zheng2024synapse,zhu2023ghost}. Voyager, SkillCraft, and Trace2Skill move further toward reusable skill packaging, while executable-skill systems such as PolySkill show how successful interaction experience can be abstracted into directly callable programmatic skills rather than free-form lessons alone~\cite{wang2023voyager,chen2026skillcraft,ni2026trace2skill,polyskill2026}. The distinction of this operation is therefore not simply that it preserves more information, but that it packages experience into artifacts that are closer to direct reuse, whether as workflow templates, structured procedures, or executable modules.

Taken together, these methods show that experience-derived acquisition is not defined by a single output form. Its unifying feature is that skills are constructed from past execution, while the main design variation lies in how those traces are processed before reuse. Different systems therefore occupy different points along a spectrum from selected traces, to summarized lessons, to structured memory, to packaged workflows and executable skill artifacts, without implying that any one form is universally superior to the others.

\subsection{Task-Derived Acquisition}

Task-derived acquisition constructs skills directly from the requirements of the current task. The task itself acts as the trigger for generation: an LLM or a related synthesis module proposes a candidate workflow, script, tool wrapper, or other skill artifact, and execution outcomes determine whether it is discarded, revised, or retained. The key idea is therefore not simply model generation, but task-conditioned skill construction followed by post-hoc retention or refinement.

For instance, CREATOR generates callable tools directly from task needs, while ToolMakers explicitly separates skill creation from skill use so that generated skills can later be reused rather than remaining tied to a single episode~\cite{qian2023creator,cai2024toolmakers}. Cradle and CodeAct synthesize procedural artifacts for immediate control and action, showing how a task can directly induce reusable operational procedures instead of only producing a final answer~\cite{tan2024cradle,wang2024codeact}. Systems such as Alita and TROVE push this pattern further by coupling task-conditioned generation with execution-time validation and later retention, whereas SkillWeaver explicitly discovers API-like skills from web interaction, filters them through subsequent use, and refines them into reusable callable artifacts over time~\cite{skillweaver2025}.

\subsection{Corpus-Derived Acquisition}

Corpus-derived acquisition derives skills from external textual or structured resources rather than from the current task or the agent's own experience. Typical sources include documentation, software repositories, datasets, interface traces, and knowledge bases, and the core process is to distill these resources into reusable procedural skills.

Representative systems illustrate different instantiations of this process. AppAgent extracts procedural signals from interface structures, AutoGuide derives context-aware guidelines from external knowledge sources, and HuggingGPT and ToolBench/ToolLLM compile procedural guidance from model cards and API descriptions. In data-science settings, DS-Agent follows a more case-driven route: it mines external competition resources (for example, Kaggle write-ups from gold- and silver-medal solutions across different types of tasks), abstracts recurring solution patterns, and turns them into reusable procedural guidance for subsequent tasks~\cite{zhang2023appagent,fu2024autoguide,shen2023hugginggpt,qin2023toolllm}.

\subsection{Discussion}

No single acquisition family dominates, and their practical importance is becoming more complementary rather than more exclusive. With the assistance of LLMs, human experts can now externalize domain knowledge into reusable skills much more easily than before, by drafting procedures, refining usage conditions, and packaging supporting resources with far lower authoring overhead. Similarly, corpus-derived acquisition has become increasingly effective because LLMs can transform documentation, repositories, case reports, and other external artifacts into more explicit procedural guidance, thereby expanding the accessible supply of candidate skills beyond what agents have personally experienced.

At the same time, the other two families remain equally important. Experience-derived acquisition is still the richest and most widely explored route, because execution traces provide a natural substrate for grounding skills in observed behavior and continuously improving them through filtering, abstraction, memory organization, and procedural packaging. Task-derived acquisition, in turn, is indispensable when agents face novel requirements that cannot wait for experts, corpora, or long-term experience accumulation, since it enables on-demand construction of candidate skills followed by validation and retention.

Taken together, these four families should not be viewed as competing alternatives, but as complementary routes for building robust skill ecosystems. Human-derived methods contribute semantic precision and high-trust expertise; experience-derived methods contribute behavioral grounding and diversity; task-derived methods contribute responsiveness; and corpus-derived methods contribute scalable cold-start coverage. In practice, the most capable skill libraries are likely to emerge from their combination, with LLMs serving as the common catalyst that lowers the cost of skill creation, transformation, and maintenance across all four routes.







\section{Skill Retrieval}
\label{sec:skill-retrieval-selection}

As agent systems accumulate reusable experience over time, the bottleneck gradually shifts from skill acquisition to skill access: the central question is no longer whether a useful skill exists, but whether the system can surface and activate the right skill at the right moment~\cite{zhao2024expel,wang2023voyager}. 
This challenge becomes especially pronounced when the skill repository grows in both scale and heterogeneity, spanning natural-language instructions, workflow memories, executable APIs, and packaged skill artifacts with metadata and auxiliary resources~\cite{zhao2024expel,wang2024agentworkflowmemory,skillweaver2025,skillnet2026}.
In such settings, skill use is better viewed as a multi-stage pipeline, where the system must first narrow the search space and then make an execution decision under the current task and environment state~\cite{wang2024agentworkflowmemory,skillrl2026}. This pipeline is often richer than one-shot semantic matching: a runtime may first discover candidate skills through names, descriptions, or metadata; progressively disclose more detailed instructions and resources only when needed; and finally bind the selected skill to concrete tools, APIs, files, or subagents for execution~\cite{zhou2026externalization}.

This perspective motivates a distinction between \textit{skill retrieval} and \textit{skill selection}. 
\textit{Skill retrieval} concerns how a large skill pool is reduced to a manageable candidate set, for example, through semantic matching, lexical lookup, generative access, or structure-aware search. 
\textit{Skill selection}, in contrast, concerns which candidate skill should ultimately be invoked, whether multiple skills should be composed, and how such choices should adapt to the current observation, subgoal, and resource budget. 
Put differently, retrieval addresses \textit{candidate recall}, while selection addresses \textit{execution-oriented decision making}. Distinguishing the two stages provides a clearer analytical framework for this section, even though practical systems often interleave them tightly.

This distinction also clarifies why retrieving skills cannot be treated as a straightforward instance of conventional document retrieval. 
Unlike documents, skills are executable units: invoking them may trigger tool calls, workflow transitions, external side effects, and non-trivial costs [2,3,4]. Consequently, semantic relatedness alone is rarely sufficient. 
A useful skill must also be executable under the current state, satisfy relevant preconditions, interact properly with other skills, and deliver sufficient utility relative to its cost~\cite{wang2024agentworkflowmemory,skillweaver2025,skillrl2026}. 
The remainder of this section therefore examines skill use as a two-stage problem: we first review how existing systems retrieve candidate skills, and then analyze how they rank, compose, and select them for execution.

\begin{figure*}[t]
\centering
\includegraphics[width=1\textwidth]{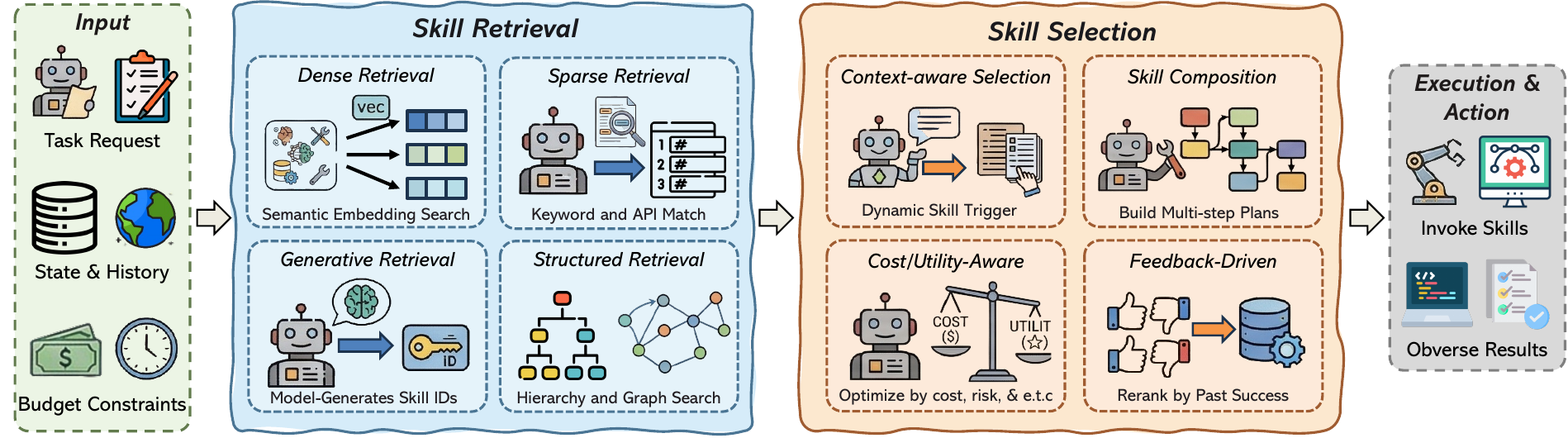}
\caption{Skill retrieval and selection.}
\label{fig:multistage-retrieval}
\end{figure*}

\subsection{Skill Retrieval}

This subsection focuses on the first stage of the skill-use pipeline: how a system retrieves a small set of candidate skills from a large repository for downstream decision making.
Unlike the introduction, the emphasis here is not on why retrieval matters in general, but on how candidate recall is operationalized in existing systems.
We organize prior work by the primary signal used to retrieve skills, including dense semantic matching, sparse lexical matching, direct generation of skill identifiers, and retrieval guided by structural relations among skills.

\subsubsection{Dense Embedding Retrieval}

Dense retrieval is the most common strategy when tasks are described in natural language and skills are paired with textual descriptions, workflow summaries, or other semantic metadata.
Its basic idea is to map the current task and candidate skills into a shared embedding space, and then retrieve the most relevant skills by vector similarity.
Voyager~\cite{wang2023voyager} remains the canonical example because semantic retrieval is applied to textual skill descriptions even though the stored skills themselves are executable code.
Recent skill-centric systems such as SAGE~\cite{wang2025reinforcement}, AutoSkill~\cite{autoskill2026}, and MemSkill~\cite{memskill2026} show the same pattern across programmatic skills, structured skill records, and explicit skills for memory management.
Lighter-weight memory-oriented variants such as ExpeL~\cite{zhao2024expel} and ReasoningBank~\cite{ouyang2025reasoningbank} extend this line by applying dense retrieval to experiential lessons or structured reasoning memories rather than fully packaged executable skills.

This makes dense retrieval the natural entry point when task formulations vary widely but the system still needs to reach reusable skills through a shared semantic layer.
The same flexibility also explains why dense retrieval is rarely the whole story.
Nearest neighbors in meaning are not always nearest neighbors in applicability, especially once libraries expose richer constraints than text similarity alone can capture~\cite{wang2025reinforcement,memskill2026}.
Accordingly, dense retrieval often opens the candidate set, while later stages refine it with metadata, structure, or execution-aware checks.

\subsubsection{Sparse and Keyword Retrieval}

Sparse retrieval treats candidate recall as matching over explicit symbolic fields and metadata attached to skill artifacts.
In the current skill literature, it operates primarily through lexical or field-aware access to those explicit descriptors.
SAGE~\cite{wang2025reinforcement} illustrates this through its ``Query N-gram'' retrieval variant, while SkillWeaver~\cite{skillweaver2025} relies heavily on explicit interface descriptions and applicability-related metadata.
AutoSkill~\cite{autoskill2026}, Memento-Skills~\cite{mementoskills2026}, and SkillNet~\cite{skillnet2026} further show that once skills are externalized as structured artifacts, lexical matching and metadata search become useful complements to semantic retrieval rather than outdated baselines.

Compared with dense retrieval, this line is narrower but often more trustworthy when the library exposes stable names, interface fields, or trigger cues.
It is therefore most useful in settings where retrieval errors come from symbolic mismatch rather than from missing semantic coverage.
Once the request becomes paraphrastic or underspecified, however, lexical evidence degrades quickly, which is why sparse retrieval usually sharpens or filters a broader candidate pool instead of replacing it~\cite{autoskill2026,mementoskills2026,skillnet2026}.

\subsubsection{Generative Retrieval}

Generative retrieval treats candidate recall as identifier generation, so the model produces the target tool or skill token directly during decoding instead of consulting a separate retrieval index.
This idea is most explicit in ToolGen~\cite{wangtoolgen}, which reformulates candidate access as identifier generation during decoding.
ToolLLM~\cite{qin2023toolllm} provides a related but looser example in which generation and large-scale tool calling are tightly coupled, even though the retrieval stage is not as explicitly reformulated.

Conceptually, this line is attractive because it removes the boundary between candidate recall and downstream action generation.
That tight coupling, however, also makes it harder to separate retrieval quality from calling behavior, and harder to guarantee coverage or identifier validity in large candidate spaces.
Within a skill survey, generative retrieval is therefore better treated as a neighboring mechanism than as the dominant pattern for explicit skill libraries.

\subsubsection{Structure-Aware Retrieval}

Structure-aware retrieval assumes that skill libraries have internal organization that should guide recall rather than treating all candidates as a flat pool.
Two recurring forms are most visible in the current literature: \textit{hierarchical retrieval}, which narrows the search space progressively, and \textit{dependency-aware retrieval}, which filters candidates through structural compatibility constraints.

\noindent $\bullet$ \textit{Hierarchical retrieval} organizes candidates through coarse-to-fine structure, so retrieval first reduces the search space at a higher level and only then resolves finer-grained choices.
SkillRL~\cite{skillrl2026} and AgentSkillOS~\cite{li2026agentskillos} are the clearest representatives because both use an explicit hierarchy to move from broad skill regions toward more specific candidates, while TOOL-PLANNER~\cite{liu2024tool} and SkillNet~\cite{skillnet2026} show the same intuition in toolkit- or ecosystem-level settings.
GraphSkill~\cite{graphskill2026} provides a related example by traversing structured documentation from high-level categories to leaf-level algorithmic entries, so coarse pruning happens before finer-grained retrieval.
This design is especially useful when the main difficulty is search-space reduction over large repositories rather than checking whether a retrieved skill is currently executable.

\noindent $\bullet$ \textit{Dependency-aware retrieval}, by contrast, focuses less on narrowing the space and more on ruling out candidates that violate prerequisites, state constraints, or composition requirements.
SkillWeaver~\cite{skillweaver2025} illustrates this through prerequisite and precondition filtering, CUA-Skill~\cite{chen2026cua} makes structural compatibility part of skill use, and ToolExpNet~\cite{zhang2025toolexpnet} shows how dependency relations can directly reshape which candidates remain plausible after retrieval.
This view becomes important once the question is not only which skill is relevant, but also which skill can participate in a valid execution path under the current conditions.


\subsubsection{discussion}

Taken together, these paradigms suggest that skill retrieval is not a single matching problem, but a trade-off among semantic flexibility, symbolic precision, and structural executability.
Dense retrieval is most effective when tasks are expressed in open-ended natural language, while sparse retrieval is more reliable when success depends on exact names, parameters, or interface-level cues.
Generative retrieval goes one step further by folding retrieval into decoding itself, but this comes with weaker control over coverage and identifier validity.
Structure-aware retrieval, in turn, is most useful when skills form a compositional space with hierarchies or dependency graphs that flat similarity cannot capture.
Overall, the field is moving away from one-shot relevance matching toward multi-signal, execution-aware candidate recall.

\subsection{Skill Selection}

Following the retrieval stage, this subsection examines how existing systems make selection decisions over the retrieved candidate set.
We organize prior work into four common perspectives: context-aware dynamic selection, skill composition, cost- and utility-aware selection, and feedback-driven re-ranking.
These categories are better understood as complementary decision dimensions than as strictly mutually exclusive classes.
A practical system may therefore combine multiple perspectives at once, such as dynamic selection with cost-aware routing or compositional planning with feedback-based updating.

\subsubsection{Context-aware dynamic selection}

Context-aware dynamic selection treats skill choice as an online decision process conditioned on the current observation, subgoal, and interaction history.
Rather than selecting a skill once from a fixed candidate set, this perspective revises the choice as execution unfolds.
This pattern is most visible when the same skill bank must support changing local states rather than only static task descriptions.
AutoGuide~\cite{fu2024autoguide} illustrates this clearly by selecting context-conditional guidelines from the current environment state.
MemSkill~\cite{memskill2026} and Memento-Skills~\cite{mementoskills2026} extend the same idea to evolving skill banks, where routing depends on the current state together with changing internal context.


\subsubsection{Skill Composition}

Skill composition views skill selection as the problem of choosing and organizing multiple skills rather than selecting a single best candidate.
Under this view, complex tasks are solved by assembling a sequence, set, or workflow of reusable skills.
The central question is therefore not only which skills are relevant, but also how they should be ordered and connected for execution.
SkillWeaver~\cite{skillweaver2025}, AWM~\cite{wang2024agentworkflowmemory}, and ASI~\cite{inducingskills2025} show this most clearly by treating APIs, workflows, or programmatic routines as reusable units that can be composed into larger behaviors.
AgentSkillOS~\cite{li2026agentskillos} and CUA-Skill~\cite{chen2026cua} push this one step further by making orchestration structure explicit through formal dependency patterns.
HuggingGPT~\cite{shen2023hugginggpt} remains a useful early reference because it established the workflow view of selection, even though it is more tool-centric than later skill-oriented systems.

This perspective anchors a large part of the current skill-selection literature because many tasks are not solved by choosing one best skill.
Instead, the system must decide how reusable modules are sequenced, grouped, or nested into a larger executable behavior.
That expressiveness comes with new failure modes, since interface compatibility, ordering constraints, and error propagation all become part of the selection problem~\cite{skillweaver2025,chen2026cua,li2026agentskillos}.
Accordingly, skill composition is usually coupled with planning or structural guidance rather than treated as simple top-k ranking.

\subsubsection{Cost \& Utility-aware selection}

The core idea of these methods is that systems should not simply prefer the most relevant skill, but should instead consider expected benefit together with cost, risk, or side effects.
MemSkill~\cite{memskill2026} and Memento-Skills~\cite{mementoskills2026} show the method side of this shift, because their routing policies are shaped by downstream utility or behavioral success rather than only surface relevance.
SkillOrchestra~\cite{wang2026skillorchestra} provides a clearer cost-aware variant that routes among candidate agents based on current skill demand, expected competence, and deployment cost.
SkillsBench~\cite{li2026skillsbench} provides the clearest empirical motivation by showing that even curated skills can have negative utility on some tasks.

This perspective matters because deployed systems pay for wrong selections not only in accuracy, but also in wasted computation, latency, and unnecessary execution.
At the same time, the literature still uses heterogeneous notions of utility, without a shared formal objective for skill selection.
So unlike the other three subclasses, cost- and utility-aware selection is best read as an emerging design criterion that should inform selection, not yet as a mature standalone family of methods.

\subsubsection{Feedback-driven reranking}

Feedback-driven re-ranking updates skill preferences using historical execution signals rather than relying only on the current candidate-query match.
The key idea is that success or failure history can be used to reorder candidates and improve later decisions.
SkillRL~\cite{skillrl2026} and CUA-Skill~\cite{chen2026cua} are the strongest skill-centric examples because execution outcomes directly alter later priorities over reusable skills.
ToolExpNet~\cite{zhang2025toolexpnet}, ExpeL~\cite{zhao2024expel}, and SMART~\cite{qian2025smart} provide adjacent variants in which experience modifies later preference ordering through feedback-informed guidance, memory revision, or calibration.

Unlike utility-aware selection, which changes what the system is trying to optimize, feedback-driven re-ranking changes how past outcomes rewrite later preference orderings.
This mechanism is most valuable for long-running agents, where today’s mistake should become tomorrow’s ranking signal.
But feedback is rarely clean, and in current systems it is often entangled with memory editing or policy adaptation rather than an isolated reranker~\cite{zhang2025toolexpnet,skillrl2026}.
For that reason, feedback-driven re-ranking usually appears as an augmentation layer over a broader selection pipeline.

\subsubsection{discussion}

Taken together, these paradigms show that skill selection is better understood as a policy problem than as a final ranking step over retrieved candidates.
Context-aware methods emphasize adaptation to the current state.
Compositional methods emphasize coordination across multiple skills.
Cost-aware methods emphasize explicit trade-offs between utility and budget.
Feedback-driven methods emphasize learning from execution outcomes over time.
These dimensions are not mutually exclusive, and many practical systems combine several of them at once.
Overall, the literature suggests that skill selection is moving from static relevance choice toward sequential, execution-aware decision making.

\begin{table*}[]
\centering
\caption{Representative methods for skill retrieval. The table compares each method by subclass, core design, structural prior or selection tags, decision inputs, and publication venue.}
\label{tab:retrieval-comparison}
\scriptsize
{\setlength{\tabcolsep}{3pt}
\renewcommand{\arraystretch}{1.18}
\resizebox{\textwidth}{!}{
\begin{tabular}{p{1.8cm} p{2.4cm} p{4.5cm} p{1.6cm} p{2.8cm} p{1.4cm}}
\toprule
\rowcolor{SkillHeaderBlue}
\textbf{SubClass} & \textbf{Method} & \textbf{Core Design} & \textbf{Tags} & \textbf{Decision Inputs} & \textbf{Venue/Year} \\
\midrule
\multicolumn{6}{c}{\textit{Skill Retrieval}} \\
\midrule
\rowcolor{SkillRowBlue}
& Voyager~\cite{wang2023voyager}  
& Text-described code skill retrieval
& Flat 
& task 
& TMLR'24 \\
\rowcolor{SkillRowBlue}
\multirow{-2}{1.8cm}{\arraybackslash Dense}
& SAGE~\cite{wang2025reinforcement}
& Program skill similarity retrieval 
& Flat 
& task instruction 
& arXiv'25 \\
\midrule
& AutoSkill~\cite{autoskill2026} 
& Hybrid rewritten-query retrieval
& Flat 
& rewrite + metadata 
& arXiv'26 \\
\multirow{-2}{1.8cm}{Sparse / \\ keyword}
& Memento-Skills~\cite{mementoskills2026} 
& Behavior-routed hybrid retrieval 
& Flat 
& state + query 
& arXiv'26 \\
\midrule
\rowcolor{SkillRowBlue}
Generative & ToolGen~\cite{wangtoolgen}   & Directly generate tool/skill identifier & Flat & task query& ICLR'25 \\
\midrule
& SkillRL~\cite{skillrl2026} 
& Hierarchical general-specific retrieval
& Hierarchy 
& instruction+skill bank
& arXiv'26 \\
\multirow{-2}{1.8cm}{Structure-aware}
& SkillWeaver~\cite{skillweaver2025}  
& Precondition-aware API filtering 
& graph 
& state + metadata & arXiv'25 \\
\midrule
\multicolumn{6}{c}{\textit{Skill Selection}} \\
\midrule
\rowcolor{SkillRowBlue}
Context-aware 
& AutoGuide~\cite{fu2024autoguide}  
& Trajectory-derived guideline routing
& Ctx 
& context + trajectory 
& NeurIPS'24 \\
\midrule
& AWM~\cite{wang2024agentworkflowmemory} 
& Workflow memory reuse
& Comp 
& task/subgoal  
& ICML'25 \\
\multirow{-2}{1.8cm}{skill\\composition}
& ASI~\cite{inducingskills2025} 
& Verified program skill composition
& Comp 
& task + prior skills 
& COLM'25 \\
\midrule
\rowcolor{SkillRowBlue}
Cost/Utility 
& MemSkill~\cite{memskill2026} 
& Reward-coupled memory routing
& Ctx+Util+Fb 
& span + memories 
& arXiv'26 \\
\midrule
& CUA-Skill~\cite{chen2026cua}  
& UI-aware failure reranking
& Ctx+Comp+Fb 
& UI + memory + failures 
& arXiv'26 \\
\multirow{-2}{1.8cm}{Feedback\\reranking}
& ToolExpNet~\cite{zhang2025toolexpnet} 
& Graph-guided experience reranking 
& Ctx+Comp+Fb 
& graph + experience 
& ACLF'25 \\
\bottomrule
\end{tabular}
}
\vspace{2pt}

\parbox{0.98\textwidth}{\footnotesize \textit{Note:} In the \textbf{Tags} column, \textit{Flat}, \textit{hierarchy}, and \textit{dependency graph} denote retrieval-side structural priors; \textit{Ctx} = context-aware, \textit{Comp} = composition, \textit{Util} = utility-aware, and \textit{Fb} = feedback-driven denote selection-side tags.}
}
\end{table*}

\subsection{Design Dimensions in Retrieval and Selection}

In this subsection, we highlight the main design dimensions that shape both stages in terms of skills retrieval and selection.

\subsubsection{Representation from the Retrieval/Selection Perspective}

Skill representation matters because retrieval and selection can only use the signals that the representation exposes.
Pure text skills mainly expose semantic and lexical cues, whereas pure code skills are often hard to retrieve or rank unless additional names, signatures, docstrings, or textual summaries are available~\cite{wang2023voyager}.
Hybrid skills expose both semantic descriptions and execution-relevant structure, which makes them easier to retrieve and easier to constrain during downstream selection~\cite{skillnet2026,skillweaver2025}.
From this perspective, representation is best understood as what the system can see and act on during retrieval and selection.

\subsubsection{State and Applicability}

State and applicability form the main bridge between retrieval and selection.
A skill may be relevant to the task description but still be unusable under the current observation, environment state, or prerequisite constraints~\cite{skillweaver2025,skillrl2026}.
State determines which candidates survive precondition or dependency checks during retrieval, and whether previously retrieved candidates remain appropriate after intermediate observations or failures during selection.

\subsubsection{Granularity and Composition}

Granularity determines what the system is actually retrieving or selecting.
Some systems operate at the level of single primitive skills, while others retrieve workflow memories, executable modules, or groups of composable skills~\cite{wang2024agentworkflowmemory,inducingskills2025}.
As granularity increases, the problem shifts from choosing one candidate to organizing a larger executable structure.
The granularity of the retrieved object therefore directly shapes whether downstream selection is a routing problem or an assembly problem.

\subsubsection{Objectives, Feedback, and Evaluation}

Objectives, feedback, and evaluation should be analyzed together because they all determine what the system is optimizing for and how that objective is updated over time.
Selection is not only about relevance, but also about whether a skill is worth invoking under constraints such as utility, reliability, or execution burden~\cite{memskill2026,mementoskills2026,li2026skillsbench}.
Execution feedback then changes future behavior by updating which candidates are preferred, suppressed, or refined~\cite{skillrl2026,mementoskills2026}.
This makes evaluation especially difficult, because standard retrieval metrics such as top-k recall do not measure whether the final execution succeeded or whether the chosen skill produced positive net utility~\cite{li2026skillsbench}.
SRA-Bench makes this gap concrete by separately examining skill retrieval, skill incorporation, and final task execution~\cite{sra2026}.
However, a complete evaluation framework still needs to connect candidate quality, execution outcome, cost efficiency, and feedback-driven adaptation rather than measuring each in isolation.

\section{Skill Evolution}
\label{sec:self-evolving-agents}

Human skills improve through correction, consolidation, and reuse rather than accumulation alone. A chess player improves when recurring mistakes reshape later pattern recognition; a craftsperson improves when failures revise an executable routine; and a clinician improves when prior cases update an earlier diagnostic heuristic. The common mechanism is \emph{progressive refinement}: feedback changes the reusable procedure behind a capability rather than merely extending a record of experience.

\begin{figure}[h]
\centering
\includegraphics[width=\columnwidth]{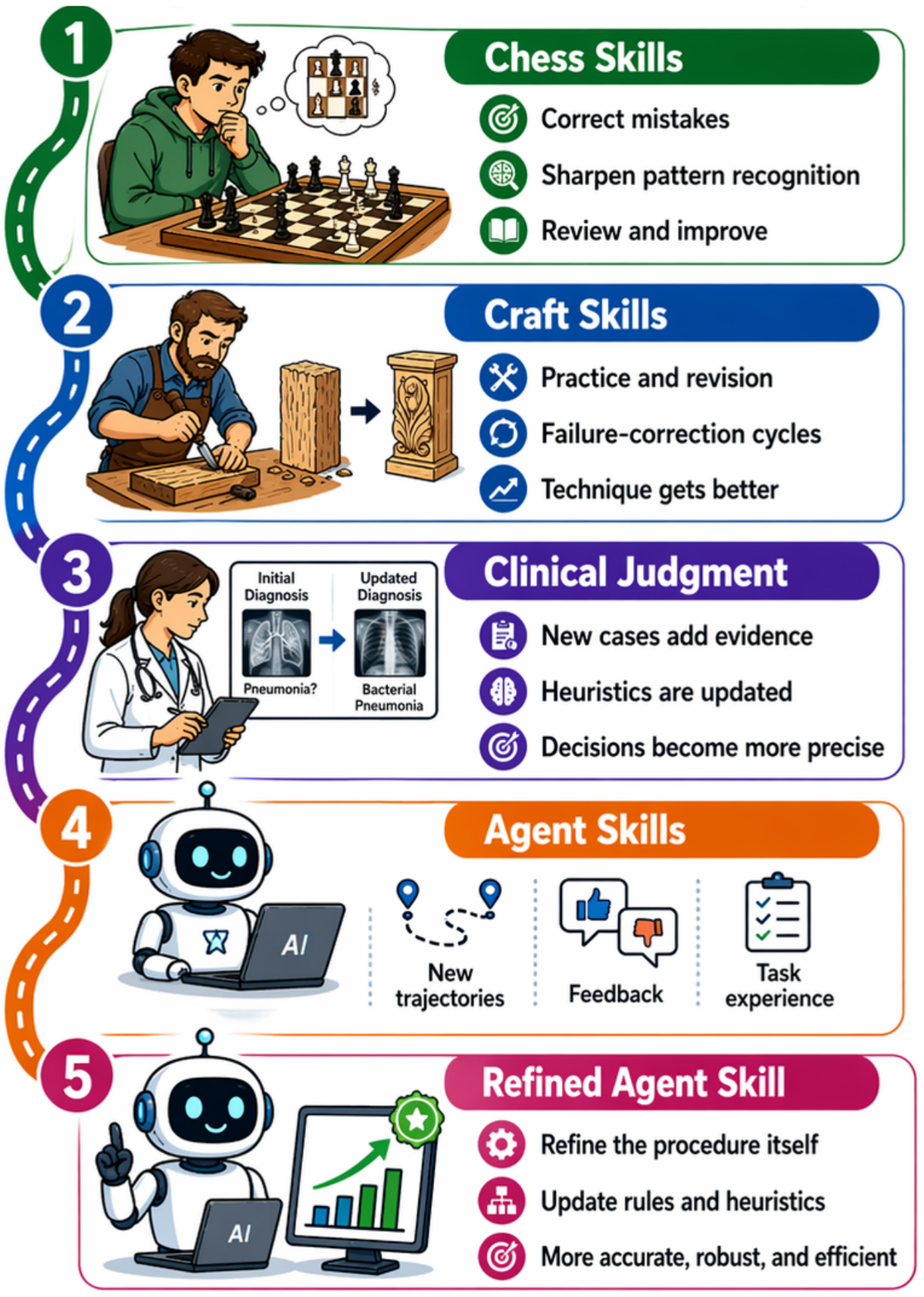}
\caption{From human skill refinement to agent skill evolution.}
\label{fig:human-agent-skill-evolution}
\end{figure}

\begin{figure*}[t]
\centering
\includegraphics[width=0.98\textwidth]{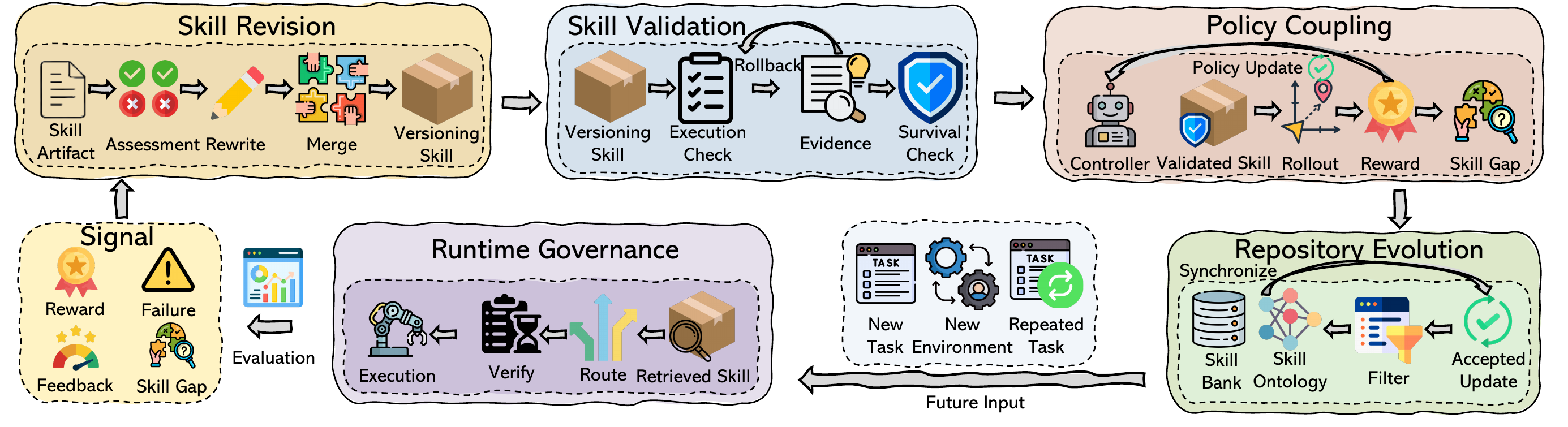}
\caption{Skill evolution through staged refinement: updates revise skills, validation filters changes, and trusted skills are indexed, retrieved, executed, and further improved.}
\label{fig:skill-evolution}
\end{figure*}

For agent systems, this analogy is useful only after skill evolution is separated from skill acquisition. Acquisition explains how a candidate skill is first obtained. Evolution asks how an already formed skill artifact is revised, validated, optimized, shared, and governed after it has been formed. Fig.~\ref{fig:human-agent-skill-evolution} provides the motivating analogy, while Fig.~\ref{fig:skill-evolution} gives the artifact-level refinement process used in the rest of this section. The evolving object may be a \texttt{Skill.md} file, a skill folder, a program API, a toolbox function, a skill bank, or a skill graph. Trajectories, task outcomes, user feedback, rewards, failures, and skill gaps are therefore update signals for the skill substrate, not the organizing categories of this section.

Skill revision changes the content of a stored artifact. Skill validation tests whether a proposed change should survive. Policy coupling lets the validated skill and the controller that uses it adapt together. Repository evolution turns accepted updates into an indexed and synchronized skill substrate. Runtime governance then retrieves candidates from that substrate under later task conditions, routes them to execution, applies trust checks, and retires unsafe or stale skills.

The cycle closes only after governed reuse. Once a governed skill is executed, its outcomes can produce the next reward, failure, feedback, skill gap, or trust signal, which returns to the revision stage. Table~\ref{tab:self-evolving-bridge-skills} compares artifact-changing methods that instantiate revision, validation, policy coupling, and repository evolution; runtime governance is discussed in prose because it controls reuse rather than directly rewriting the skill artifact.

\begingroup
\definecolor{SkillHeaderBlue}{RGB}{222,232,245}
\definecolor{SkillRowBlue}{RGB}{241,246,253}
\begin{table*}[t]
\centering
\caption{Representative methods that instantiate stages of skill refinement. Each row identifies the durable skill-side artifact, the trigger that updates it, the evolution operator, the validation check that controls survival, and the scope over which the evolved skill is reused.}
\label{tab:self-evolving-bridge-skills}
\scriptsize
{\setlength{\tabcolsep}{2.4pt}
\renewcommand{\arraystretch}{1.08}
\resizebox{\textwidth}{!}{
\begin{tabular}{
>{\raggedright\arraybackslash}p{2.25cm}
>{\raggedright\arraybackslash}p{2.7cm}
>{\raggedright\arraybackslash}p{2.65cm}
>{\raggedright\arraybackslash}p{3.05cm}
>{\raggedright\arraybackslash}p{3.25cm}
>{\raggedright\arraybackslash}p{1.6cm}}
\toprule
\rowcolor{SkillHeaderBlue}
\textbf{Method} & \textbf{Evolving Artifact} & \textbf{Update Trigger} & \textbf{Evolution Operator} & \textbf{Validation Check} & \textbf{Reuse Scope} \\
\midrule
\multicolumn{6}{c}{\textit{Skill Revision}} \\
\midrule
\rowcolor{SkillRowBlue}
EvoSkill~\cite{evoskill2026} & skill folder & failure trace & create, rewrite & held-out validation & cross-task \\
Memento-Skills~\cite{mementoskills2026} & skill folder & execution feedback & attribute, rewrite & unit-test rollback & cross-session \\
\rowcolor{SkillRowBlue}
AutoSkill~\cite{autoskill2026} & \texttt{SKILL.md} & user feedback + dialogue history & add, merge, discard & maintainer approval & personal \\
XSkill~\cite{xskill2026} & skill document + experience records & visual rollout + usage history & update, merge, remove, refine & skill-manager check & cross-task \\
\midrule
\multicolumn{6}{c}{\textit{Skill Validation}} \\
\midrule
\rowcolor{SkillRowBlue}
SkillWeaver~\cite{skillweaver2025} & program API & practice outcome & debug, document, update & generated tests + debugging & cross-agent \\
ASI~\cite{inducingskills2025} & program skill & successful trajectories, website changes & induce, verify, update & test-trajectory execution & cross-task \\
\rowcolor{SkillRowBlue}
CoEvoSkills~\cite{coevoskills2026} & multi-file skill package & surrogate-verifier feedback & generate, refine, co-evolve & surrogate verification & cross-task \\
TroVE~\cite{trove2024} & function toolbox & streaming executions & grow, select, trim & execution agreement & cross-task \\
\rowcolor{SkillRowBlue}
PSN~\cite{shi2026psn} & skill graph & failure trace & repair, refactor & maturity gate + rollback validation & cross-task \\
Audited-SG~\cite{auditedskillgraph2025} & skill graph & verifier report & verify, promote & replayable evidence bundle & ecosystem \\
\midrule
\multicolumn{6}{c}{\textit{Policy Coupling}} \\
\midrule
\rowcolor{SkillRowBlue}
SkillRL~\cite{skillrl2026} & SkillBank & RL reward + validation failures & generate, refine SkillBank & validation checkpoint & cross-task \\
ARISE~\cite{arise2026} & tiered skill library & successful traces + hierarchical reward & select, generate, add, update, evict & confidence gate + skill validation & cross-task \\
\midrule
\multicolumn{6}{c}{\textit{Repository Evolution}} \\
\midrule
\rowcolor{SkillRowBlue}
Uni-Skill~\cite{uniskill2026} & SkillFolder repository & skill gap & expand, ground missing skills & video-aligned examples & cross-task \\
SkillX~\cite{skillx2026} & multi-level skill KB & execution feedback + exploration trace & refine, expand, update & skill filter & cross-task \\
\rowcolor{SkillRowBlue}
SkillNet~\cite{skillnet2026} & skill repository + ontology & repository contribution & create, evaluate, connect & repository rubric & ecosystem \\
SkillClaw~\cite{skillclaw2026} & shared skill repository & cross-user trajectory & refine, create, synchronize & environment validation + synchronization check & cross-user \\
\bottomrule
\end{tabular}
}
\vspace{2pt}

\parbox{0.98\textwidth}{\footnotesize \textit{Note:} The table excludes experience-only, memory-only, one-shot acquisition, routing-only, and threat-analysis-only methods. \textbf{Update Trigger} denotes the signal that changes an existing skill substrate; \textbf{Evolution Operator} denotes the artifact-level operation; \textbf{Validation Check} denotes the evidence used to control survival, reuse, or ecosystem acceptance.}
}
\end{table*}
\endgroup

\subsection{Skill Revision}
\label{subsec:self-evolving-skill-revision}

Skill revision is the content-changing stage of evolution: feedback modifies a persistent skill object, and the system determines whether the modification should survive. EvoSkill~\cite{evoskill2026} fits this pattern because failed executions trigger a decision about whether to refine an existing skill or create a missing one. The accepted result is materialized as a structured skill folder with trigger metadata, instructions, and helper scripts. The held-out validation step is what makes this evolution rather than unconstrained self-editing: a candidate folder must improve future performance before it is retained.

Deployment-time revision adds a stronger requirement: the skill must be rewritten without corrupting the future library. Memento-Skills~\cite{mementoskills2026} addresses this by reading a relevant skill folder, executing under that guidance, and attributing failure before rewriting the prompt or program that will be reused later. Its unit-test gate and rollback step make revision reversible. The update is therefore not merely memory read and write, but a sequence of execution, attribution, rewriting, validation, and later reuse.

Longitudinal revision emphasizes identity across updates. AutoSkill~\cite{autoskill2026} represents recurring behaviors as editable \texttt{SKILL.md} artifacts and updates them through add, merge, or discard decisions when user feedback refines an existing capability. XSkill~\cite{xskill2026} extends this idea to multimodal agents by maintaining a skill document alongside experience records; its skill manager can update, merge, remove, or refine skill content as visually grounded rollouts and usage history accumulate. In both cases, the evolving unit is not a new trace but a persistent skill whose content is consolidated across later use.

The common claim is therefore narrow but strong. These systems demonstrate skill evolution when feedback changes a named, reusable artifact and when that artifact remains available after the update. They do not require the survey to treat every trajectory-driven skill construction method as evolution; the evidence lies in artifact revision, versioning, validation, and later reuse.

\subsection{Skill Validation}
\label{subsec:self-evolving-programmatic-verification}

Skill validation turns evolution into a survival problem: a revised skill must pass a check before it is trusted as future capability. SkillWeaver~\cite{skillweaver2025} constructs web-agent skills as APIs and hones them through practice, generated tests, and documentation updates---useful evidence, though weaker than full semantic verification since passing tests may only show that a function avoids immediate execution failure. Agent Skill Induction (ASI)~\cite{inducingskills2025} sharpens this check by treating executability as the boundary between acquisition and evolution: induced programs enter the refinement cycle only when verified against test trajectories, and prior skills can be updated when websites change. CoEvoSkills~\cite{coevoskills2026} targets richer multi-file skill packages and couples a skill generator with a surrogate verifier, so verification feedback itself co-evolves with the skill without access to ground-truth test content. TroVE~\cite{trove2024} extends the unit of repair from one skill to a function toolbox, growing and trimming entries via execution agreement, while Programmatic Skill Networks (PSN)~\cite{shi2026psn} push further to a graph of executable symbolic skills with failure localization, maturity gates, and rollback-validated refactoring. Audited Skill-Graph~\cite{auditedskillgraph2025} adds an evidence layer by promoting candidate skills into a directed graph only when supported by verifier reports and replayable evidence bundles. Across these methods, executable skill evolution depends on a survival check---execution, agreement, replay, or rollback---before a proposed update becomes part of the reusable substrate.

\subsection{Policy Coupling}
\label{subsec:self-evolving-policy-optimization}

Policy coupling treats the skill substrate as part of the controller's training state. SkillRL~\cite{skillrl2026} builds a hierarchical SkillBank, retrieves general and task-specific skills during rollouts, and performs recursive skill evolution during reinforcement learning. Validation failures expose regions where existing skills are insufficient, while reward-driven policy updates change which failures and skill gaps the system encounters next. The skill bank is therefore a dynamic training component rather than a static context cache.

ARISE~\cite{arise2026} makes the coupling explicit through a hierarchical manager--worker design. The skills manager selects relevant skills before execution and summarizes successful traces into a tiered cache--reservoir library after execution. Hierarchical rewards encourage both task success and useful skill use, while confidence and validation checks control whether generated or updated skills are inserted. The library is part of the agent state: selection, generation, addition, update, eviction, loading, and deletion co-adapt with the policy.

This stage is narrower than general continual learning. A method belongs here only when policy optimization changes the skill substrate and the changed substrate later affects the policy's action space or rollout distribution. This is why CASCADE~\cite{cascade2025} is treated later as a boundary case: it provides useful evidence about cumulative executable scientific routines, but its central mechanism is continuous learning and self-reflection for scientific tool use rather than RL-style co-optimization of a skill library and controller.

\subsection{Repository Evolution}
\label{subsec:self-evolving-repository}

Repository evolution asks how accepted skill changes scale beyond a single artifact. Uni-Skill~\cite{uniskill2026} targets robotic manipulation settings where a fixed library cannot cover novel task decompositions. Its planner detects missing skills, requests supplementary skill descriptions, and uses SkillFolder to ground those descriptions in a hierarchical repository of annotated robotic video segments. The evolution claim should therefore be read as repository expansion and grounding, not as mature autonomous revision of existing entries.

The case for repository evolution is stronger when the repository itself is refined, filtered, and connected. SkillX~\cite{skillx2026} treats a multi-level skill knowledge base as the object being improved: trajectories are distilled into planning, functional, and atomic skill levels; the resulting skills are refined and filtered; and coverage expands through experience-guided exploration. SkillNet~\cite{skillnet2026} shifts the emphasis to shared infrastructure. It organizes a large repository through dynamic ontology construction, relation graphs, data-driven filtering, and multi-dimensional evaluation covering safety, completeness, executability, maintainability, and cost-awareness.

Collective evolution adds a synchronization problem. SkillClaw~\cite{skillclaw2026} aggregates trajectories across users, lets an agentic evolver refine existing skills or create new ones, validates candidate updates in user environments, and synchronizes accepted changes back to the shared repository. This makes repository evolution depend on both update quality and propagation control. Local systems must avoid bad rewrites; repository systems must also avoid duplicate skills, inconsistent relations, weak coverage, and unsafe distribution. The output of this stage is therefore not an executed skill, but a searchable substrate from which runtime governance can retrieve candidates under a later task context.

\subsection{Runtime Governance}
\label{subsec:self-evolving-routing-governance}

An evolved skill changes behavior only if runtime control retrieves and uses it under the right conditions. SkillRouter~\cite{skillrouter2026} is therefore not a core revision method, but it exposes a necessary selection layer for large skill pools. Its retrieve-and-rerank pipeline shows that the full skill body provides stronger routing evidence than the name or description alone. Once a skill has been refined and indexed, the system still needs a selector that can retrieve the changed artifact rather than an obsolete or superficially similar one.

Governance addresses the corresponding failure mode: an evolved or shared skill may be executable, but not safe to trust at runtime. Audited Skill-Graph~\cite{auditedskillgraph2025} represents the positive side of this boundary by tying promotion to replayable evidence. PoisonedSkills~\cite{poisonedskills2026} represents the negative side: third-party skill documentation can hide malicious logic that agents later execute as trusted operational guidance. SkillClaw~\cite{skillclaw2026} similarly shows that collective evolution requires validation before synchronized updates are propagated to users. 

Runtime governance closes the staged cycle by deciding whether a stored update actually affects future behavior. Revision changes a skill, validation decides whether the change should survive, and policy or repository mechanisms propagate the result. Runtime governance then retrieves, routes, trust-checks, executes, or retires the candidate. Execution produces the next reward, failure, feedback, skill gap, or trust signal. Provenance, permission boundaries, contamination detection, and retirement are needed so that an evolving ecosystem does not simply accumulate more executable risk.

\subsection{Discussion}
\label{subsec:self-evolving-discussion}

The evidence for skill evolution is strongest when a method directly modifies a named artifact and exposes a check on the update---revision methods such as EvoSkill~\cite{evoskill2026}, Memento-Skills~\cite{mementoskills2026}, AutoSkill~\cite{autoskill2026}, and XSkill~\cite{xskill2026}; validation methods such as SkillWeaver~\cite{skillweaver2025}, ASI~\cite{inducingskills2025}, CoEvoSkills~\cite{coevoskills2026}, TroVE~\cite{trove2024}, and PSN~\cite{shi2026psn}; and substrate-level methods such as SkillRL~\cite{skillrl2026}, ARISE~\cite{arise2026}, SkillX~\cite{skillx2026}, and SkillClaw~\cite{skillclaw2026}---whose common denominator is that a durable substrate is changed and later reused under a survival condition. Repository-scale autonomy is more partial: Uni-Skill~\cite{uniskill2026}, SkillNet~\cite{skillnet2026}, and Audited Skill-Graph~\cite{auditedskillgraph2025} make repository evolution inspectable and governable but do not settle long-horizon causal attribution once many skills, routers, evaluators, and users interact. Boundary cases sharpen the scope: CASCADE~\cite{cascade2025} demonstrates cumulative skill growth without artifact-level revision; SkillRouter~\cite{skillrouter2026} and PoisonedSkills~\cite{poisonedskills2026} show that later selection and trust can determine whether evolution helps or harms deployment; and SKILL0~\cite{skillzero2026} trades external governance for runtime efficiency by internalizing skills through an in-context RL curriculum. 

\section{Open Challenges}
\label{sec:challenges}

Despite the well-established lifecycle of agent skills, several critical challenges remain open across its various stages.

\subsection{Skill Acquisition}

\textbf{Abstraction quality.} 
Experience-derived acquisition must decide what to preserve from noisy trajectories and what to discard. Skills that remain too local behave like episodic memories; skills that over-abstract lose operational value.

\textbf{Weak trigger specification.} 
Many acquisition pipelines produce a plausible procedure but weak conditions for when it should be used. As a result, the skill itself may be useful, yet still fail in deployment because it is routed poorly.

\textbf{Resource drift.} 
As libraries mature, attached scripts, references, and schemas can become stale or inconsistent with the main document. Acquisition is therefore not a one-shot step---it creates artifacts that later sections must retrieve, evolve, and maintain.

\textbf{Admission quality at scale.} 
Faster acquisition pipelines can generate candidate skills more quickly than libraries can validate and curate them. This creates a familiar form of technical debt: low-quality skills accumulate, retrieval becomes noisier, and orchestration quality deteriorates.

\subsection{Skill Retrieval}

The current literature already offers several retrieval and selection strategies, but the central difficulties now lie less in defining isolated method categories than in making skill use robust at scale.
These open challenges arise because skills are executable, structured, and continuously evolving artifacts rather than passive knowledge snippets.

\textbf{Scalable skill libraries.}
Retrieval and selection must remain robust as skills are added, merged, revised, or deprecated over time. Existing systems make this challenge apparent, yet they still lack a general mechanism for keeping indexes, metadata, and priorities consistently synchronized as the skill library evolves~\cite{autoskill2026,skillnet2026,li2026agentskillos,mementoskills2026}.

\textbf{Constraint-aware composition.}
The problem is not only to find relevant skills, but to find skills that can be connected into a feasible execution path.
Current work handles pieces of this through prerequisites, workflow memories, or compositional modules, but these constraints are still modeled in system-specific ways rather than through a shared abstraction~\cite{skillweaver2025,skillrl2026,chen2026cua,wang2024agentworkflowmemory,inducingskills2025}.

\textbf{Multi-objective selection.}
Behavioral success and utility can diverge from surface relevance, yet the field still lacks a common way to combine success, cost, latency, safety, risk, and user preference into one decision objective~\cite{memskill2026,mementoskills2026,li2026skillsbench}.
Many systems therefore still optimize a narrow proxy while leaving the rest of the decision problem implicit.

\textbf{Execution-centric evaluation.}
Retrieval recall or selection accuracy cannot show whether the final decision improved end-to-end execution, saved cost, or introduced negative downstream effects~\cite{li2026skillsbench}.
Stronger evaluations should connect candidate quality, final decision quality, execution success, efficiency, recovery, and long-term utility in a single protocol.

\textbf{Personalized selection.}
The same task can call for different skills under different user histories, habits, constraints, or preferences.
Early personalized skill-bank systems suggest the promise of this direction, but personalization has not yet become a first-class part of the selection objective~\cite{autoskill2026}.

\textbf{Adaptive selection and recovery.}
In real environments, skill use is a continuing control process: agents must revise choices after failures, switch to substitutes, and replan under updated states~\cite{skillrl2026,memskill2026,mementoskills2026}. Thus, retrieval, selection, execution, and library evolution should be jointly modeled rather than treated as isolated steps. Future progress will rely not only on better similarity matching, but also on scalable skill management, structured constraints, richer objectives, and execution-centered evaluation.

\subsection{Skill Evolution}

\textbf{Coarse artifact-level evaluation.}
Final task success cannot reveal whether a promoted artifact was actually reused, transferred across tasks, or selected for the right reason. TRACE~\cite{trace2025} and STULIFE~\cite{stulife2025} push toward lifecycle behavior rather than endpoint accuracy, while MIRAGE~\cite{mirage2025} warns that correct answers do not establish that the underlying mechanism was appropriate. Most of the literature still reports outcome gains without isolating the causal contribution of the evolving artifact.

\textbf{Asymmetric revision.}
Current systems are far better at adding artifacts than at safely rewriting or retiring them. EvolveR~\cite{evolver2025} explicitly scores, merges, and prunes principles; Memento-Skills~\cite{mementoskills2026} makes rollback and write gating unavoidable once the skill library is writable; MemEvolve~\cite{memevolve2025} shows the asymmetry one layer deeper, where better content cannot compensate for an inadequate container. Growth is better understood than cleanup.

\textbf{Weakly specified repository-scale governance.}
Once artifacts become shareable, the question shifts from \emph{what to retain} to who can publish, trust, deprecate, and bear responsibility when an artifact is wrong. SkillNet~\cite{skillnet2026} and SkillOS~\cite{skillos2026} address asset management; SkillRouter~\cite{skillrouter2026} shows that routing is itself a control bottleneck; skills.vote~\cite{skillsvote2026}, Audited Skill-Graph~\cite{auditedskillgraph2025}, PoisonedSkills~\cite{poisonedskills2026}, and Anthropic Agent Skills~\cite{anthropicagentskills2025} expose the pressures of discoverability, provenance, contamination, and externally managed ecosystem trust.

\textbf{Confounded gains and long-horizon trust.}
Reported improvements may stem not from better artifacts but from stronger judges, better task synthesis, richer grounding, or more test-time compute---as visible in AgentEvolver~\cite{agentevolver2025}, Explorer~\cite{explorer2025}, and ReasoningBank~\cite{ouyang2025reasoningbank}, with the UI data-scale study~\cite{uidatascale2024} showing that scale alone can move metrics substantially. Writable memories, bridge objects, and skill repositories further introduce long-horizon trust requirements that current evaluations only partly address~\cite{wang2024agentworkflowmemory,ace2025,park2023generative,memoryagents2026,rethinkingmemory2026,sokagenticskills2026}. Robust self-evolution therefore depends on trustworthy lineage, interpretable revision, and causal evaluation, not on persistence and reuse alone.
\section{Future Research Directions} \label{sec:future}

We outline five directions for advancing agent skill research.

\textbf{Unified Skill Schema.} Despite broad adoption of the skill abstraction, there is substantial heterogeneity in how $\mathcal{M}$ is written and how auxiliary resources are linked and maintained. A standardized schema defining common fields for scope, triggering conditions, dependencies, versioning, and safety constraints would make skills easier to share, retrieve, and govern across ecosystems.

\textbf{Resource-Aware Joint Optimization.} Current systems treat retrieval, planning, and execution as isolated stages~\cite{lewis2020rag,react2022}. At scale, inference latency, token costs, and tool invocation risk become first-class constraints~\cite{cai2024toolmakers}. We should pursue end-to-end optimization that navigates the trade-off between utility, latency, and execution cost, including dynamic model routing and workload-aware scheduling~\cite{wu2023autogen,hong2024metagpt}.

\textbf{Skill Library Evolution under Non-Stationarity.} APIs deprecate, tool behavior shifts, and task distributions change over time~\cite{qin2023toolllm,anthropic2024mcp}. Skill libraries need lifecycle-level robustness: drift detection, compatibility checks, safe online updates, and versioned rollback. Benchmarks should measure post-deployment stability and regression recovery, not only zero-shot success~\cite{shen2024taskbench}.

\textbf{Multimodal and Domain-Specific Benchmarks.} Current benchmarks are concentrated in text-centric settings. Practical deployments increasingly require perception-action integration under domain constraints. Embodied agents, autonomous driving, and low-altitude UAV scenarios are representative targets, where skills must be evaluated for safety, latency, and long-horizon decision quality.

\textbf{Causality-Driven Skill Diagnosis.} Many failures span retrieval, selection, and execution and cannot be resolved from surface-level logs~\cite{xia2024agentless,shinn2023reflexion}. Systems should attribute failures to specific causes—retrieval mismatch, policy mis-selection, tool malfunction, or unsafe composition—and trigger targeted repair. This requires trace-level causal telemetry that connects runtime decisions to post-hoc diagnosis and safe updates~\cite{wu2023autogen}.
\section{Application Scenarios}

\begin{figure}[t]
\centering
\includegraphics[width=1.0\columnwidth]{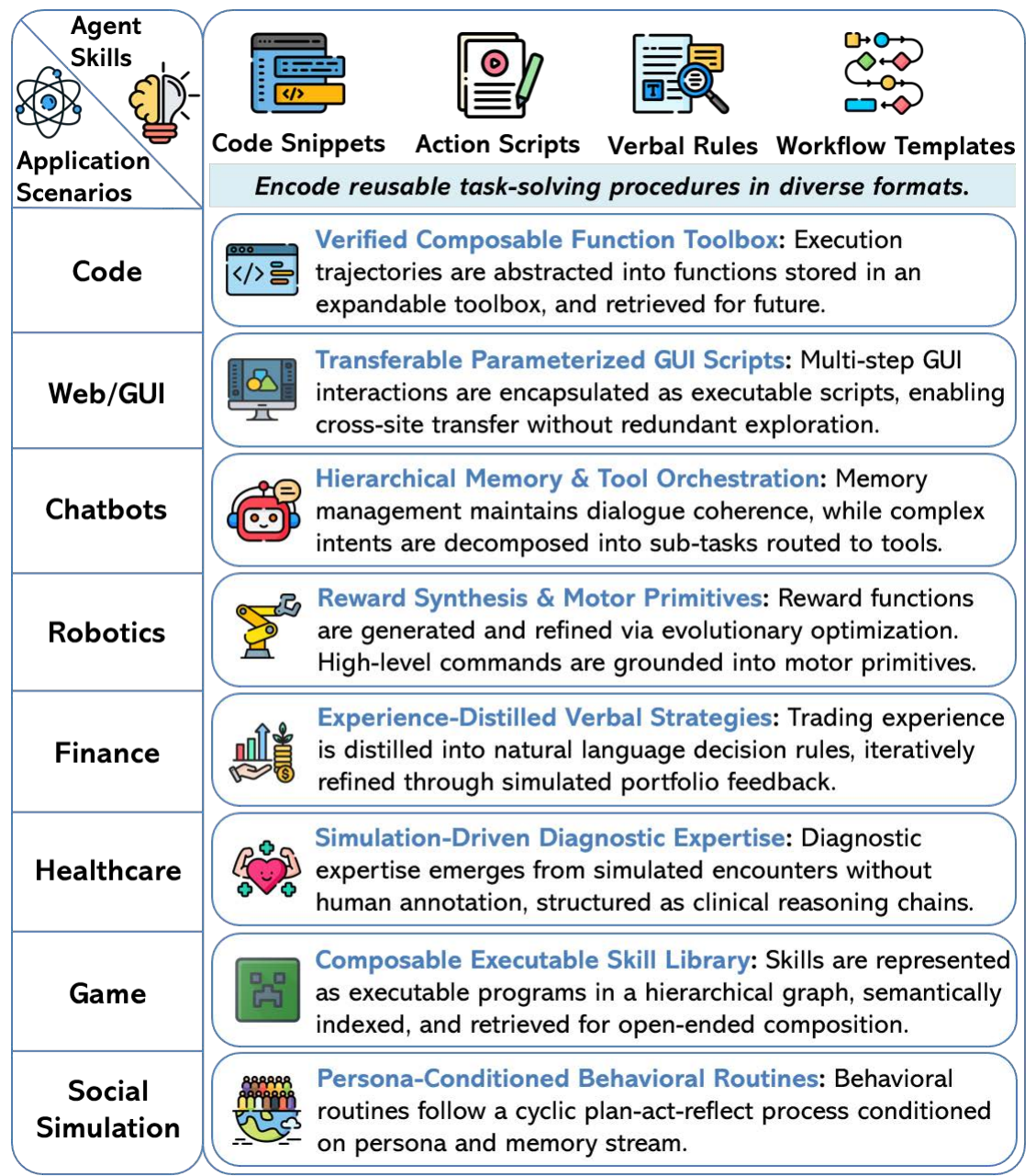}
\caption{Application scenarios of agent skills.}
\label{fig:application_scenarios}
\end{figure}

Agent skills manifest differently depending on domain demands. We organize representative scenarios into eight categories and highlight dominant skill formats and acquisition strategies. Fig. \ref{fig:application_scenarios} illustrates these scenarios.

In software engineering, skills package recurring workflows such as code generation, debugging, and repository-aware refactoring so agents can apply verified routines instead of solving each issue from scratch~\cite{trove2024,wang2024codeact,yang2024sweagent,zhong2024ldb,yue2024dsagent,zhang2023toolcoder}. In web and GUI tasks, skills encode multi-step interaction routines over dynamic interfaces, supporting navigation, form completion, and recovery from UI changes~\cite{skillweaver2025,chen2026cua,zhang2023appagent,zheng2024synapse,wang2024agentworkflowmemory,zhou2024webarena}. In chatbot systems, skills stabilize long-horizon dialogue by encoding memory-update policies, tool-routing routines, and failure-recovery procedures~\cite{packer2023memgpt,wang2025mplus,liu2023thinkinmemory,nemori2025,yang2026plugmem,shen2023hugginggpt}. In robotics, skills serve as reusable control routines connecting perception, action, and reward optimization, enabling agents to compose and adapt motor behaviors across tasks~\cite{saycan2022,ma2023eureka,polyskill2026,uniskill2026}. In finance, skills encode reusable decision heuristics for analysis, planning, and portfolio adjustment under changing market conditions~\cite{yu2024fincon}. In healthcare, skills structure diagnostic reasoning and treatment planning into consistent procedures for medical decision support~\cite{li2024agenthospital}. In game environments, skills are composable behavioral units discovered through continual interaction, enabling agents to build and reuse complex action repertoires during long-horizon tasks~\cite{wang2023voyager,zhu2023ghost,deps2023,wang2023jarvis1}. In social simulation, skills encode reusable social-behavior routines that coordinate planning, acting, and reflection under  constraints~\cite{park2023generative,damcs2025}.


\section{Related Work}
\label{sec:related}

This section reviews neighboring research streams and clarifies how our survey positions itself relative to prior literature.

$\bullet$ \textbf{Tool Use in LLM Agents.}
A broad line of work studies how LLMs invoke external tools, APIs, databases, and execution environments to improve factuality and actionability~\cite{schick2023toolformer,react2022,qin2023toolllm,shen2023hugginggpt}. Early systems are largely API-centric, focusing on call planning and execution in prompted reasoning-action loops~\cite{react2022,wang2023planandsolve,yao2023tree,shinn2023reflexion}. As tool spaces expand, \emph{tool retrieval} becomes a first-class problem: systems must map open-ended user intents to the right tool among large candidate sets, with growing attention to query-tool alignment, retrieval quality, and retrieve-then-call integration~\cite{li2023api,shi2025retrieval,zhang2024data,wangtoolgen,liutoolace,yuan2025easytool}. 
The infrastructure narrative then shifts from ad hoc API wiring to protocol-guided interoperability: function-calling schemas and MCP-style interfaces separate reasoning from capability serving, enabling standardized access control, schema mediation, and cross-provider composition~\cite{openai2023functioncalling,anthropic2024mcp,langchain2022,zhou2026externalization}.

$\bullet$ \textbf{RAG and Agent Memory.}
Retrieval-centric research provides core mechanisms for non-parametric grounding, from dense/sparse retrieval and classic RAG pipelines to agentic retrieval under task constraints~\cite{karpukhin2020dpr,lewis2020rag,du2024anytool,ferraz2026exprag}. More recent lines also move beyond flat text retrieval toward structure-aware and graph-based retrieval settings, where document hierarchy, dependency structure, or graph organization influences candidate construction and reranking~\cite{graphskill2026,damcs2025,zhou2025graphbasedrag,su2025cluerag,zhang2025erarag}. In parallel, memory-oriented work studies how agents preserve and reactivate long-horizon experience through memory management and consolidation~\cite{packer2023memgpt,liu2023thinkinmemory,nemori2025,intrinsicmemory2025}. 
System-level memory architectures and recent long-term memory systems organize experience for multi-step agent coordination and extended-context use~\cite{zhang2506g,damcs2025,hu2026evermemos,yue2026hypermem,chen2026msa}. Complementary benchmarks evaluate such memory capabilities through test-time adaptation and memory quality~\cite{evomemory2025}.
Although closely related and discussed in externalization-oriented syntheses~\cite{zhou2026externalization}, these works center on retrieval and memory state, whereas our survey examines reusable procedural skills and their lifecycle in agent ecosystems.
\section{Conclusion}
\label{sec:conclusion}

This survey examines LLM-based agent systems through the lens of agent skills—reusable procedural artifacts that coordinate tools, memory, and runtime context. While agents handle high-level reasoning and planning, skills from the operational layer enable reliable, composable execution. We organize the literature around four lifecycle stages: representation, acquisition, retrieval and selection, and evolution, offering a unified account of how procedural knowledge is externalized, activated, and maintained. Treating skills as first-class building blocks, rather than incidental prompts or tool wrappers, is key to improving the scalability, robustness, and governability of agent systems. We hope this framework provides a useful foundation for future research on agent skill systems, their supporting infrastructures, and the long-term management of reusable machine capabilities.

		

\bibliographystyle{IEEEtran}
\bibliography{agent_survey}

\end{document}